%Paper: hep-th/9207094
%From: WITTEN%IASSNS.BITNET@pucc.princeton.edu
%Date: Tue, 28 Jul 92 15:45 EST
%%%%%%%%%%%%%%%%%%%%%%%%REVISED 12/19/03
%
% % % % % % % % % % % % % % % % % % % % % % % % % % % % % % % % % % % %
%%  Inserted Figures.
%
%
\newbox\picturebox
\def\p@cht{\ht\picturebox }
\def\p@cwd{\wd\picturebox }
\def\p@cdp{\dp\picturebox }
\newdimen\xshift
\newdimen\yshift
\newdimen\captionwidth
\newskip\captionskip
\captionskip=15pt plus 5pt minus 3pt
\def\fullwidth{\captionwidth=\hsize }
\newtoks\Caption
\newif\ifcaptioned
\newif\ifselfcaptioned
\def\caption{\captionedtrue \Caption }
\newcount\linesabove
\newif\iffileexists
\newtoks\picfilename
\def\fil@#1 {\fileexiststrue \picfilename={#1}}
\def\file#1{\if=#1\let\n@xt=\fil@ \else \def\n@xt{\fil@ #1}\fi \n@xt }
\def\pl@t{\begingroup \pr@tect
    \setbox\picturebox=\hbox{}\fileexistsfalse
    \let\height=\p@cht \let\width=\p@cwd \let\depth=\p@cdp
    \xshift=\z@ \yshift=\z@ \captionwidth=\z@
    \Caption={}\captionedfalse
    \linesabove =0 \picturedefault }
\def\plot{\pl@t \selfcaptionedfalse }
\def\Picture#1{\gl@bal\advance\figurecount by 1
    \xdef#1{\the\figurecount}\pl@t \selfcaptionedtrue }

\def\s@vepicture{\iffileexists \parsefilename \redopicturebox \fi
   \ifdim\captionwidth>\z@ \else \captionwidth=\p@cwd \fi
   \xdef\lastpicture{\iffileexists
        \setbox0=\hbox{\raise\the\yshift \vbox{%
              \moveright\the\xshift\hbox{\picturedefinition}}}%
        \else \setbox0=\hbox{}\fi
         \ht0=\the\p@cht \wd0=\the\p@cwd \dp0=\the\p@cdp
         \vbox{\hsize=\the\captionwidth \line{\hss\box0 \hss }%
              \ifcaptioned \vskip\the\captionskip \noexpand\Tenpoint
                \ifselfcaptioned Figure~\the\figurecount.\enspace \fi
                \the\Caption \fi }}%
    \endgroup }
\let\endpicture=\s@vepicture
\def\savepicture#1{\s@vepicture \global\let#1=\lastpicture }
\def\displaypicture{\fullwidth \s@vepicture $$\lastpicture $${}}
\def\toppicture{\fullwidth \s@vepicture \topinsert
    \lastpicture \medskip \endinsert }
\def\midpicture{\fullwidth \s@vepicture \midinsert
    \lastpicture \endinsert }
%
%  Wraparound macros - a try.
%
\def\leftpicture{\pres@tpicture
    \dimen@i=\hsize \advance\dimen@i by -\dimen@ii
    \setbox\picturebox=\hbox to \hsize {\box0 \hss }%
    \wr@paround }
\def\rightpicture{\pres@tpicture
    \dimen@i=\z@
    \setbox\picturebox=\hbox to \hsize {\hss \box0 }%
    \wr@paround }
\def\pres@tpicture{\gl@bal\linesabove=\linesabove
    \s@vepicture \setbox\picturebox=\vbox{
         \kern \linesabove\baselineskip \kern 0.3\baselineskip
         \lastpicture \kern 0.3\baselineskip }%
    \dimen@=\p@cht \dimen@i=\dimen@
    \advance\dimen@i by \pagetotal
    \par \ifdim\dimen@i>\pagegoal \vfil\break \fi
    \dimen@ii=\hsize
    \advance\dimen@ii by -\parindent \advance\dimen@ii by -\p@cwd
    \setbox0=\vbox to\z@{\kern-\baselineskip \unvbox\picturebox \vss }}
\def\wr@paround{\Caption={}\count255=1
    \loop \ifnum \linesabove >0
         \advance\linesabove by -1 \advance\count255 by 1
         \advance\dimen@ by -\baselineskip
         \expandafter\Caption \expandafter{\the\Caption \z@ \hsize }%
      \repeat
    \loop \ifdim \dimen@ >\z@
         \advance\count255 by 1 \advance\dimen@ by -\baselineskip
         \expandafter\Caption \expandafter{%
             \the\Caption \dimen@i \dimen@ii }%
      \repeat
    \edef\n@xt{\parshape=\the\count255 \the\Caption \z@ \hsize }%
    \par\noindent \n@xt \strut \vadjust{\box\picturebox }}
\let\picturedefault=\relax
\let\parsefilename=\relax
\def\redopicturebox{\let\picturedefinition=\rel@x
   \errhelp=\disabledpictures
   \errmessage{This version of TeX cannot handle pictures.  Sorry.}}
\newhelp\disabledpictures
     {You will get a blank box in place of your picture.}
%
%
%
% Above definitions of \parsefilename and \redopicturebox
% are dumb defaults.  Actual definition are system dependent;
% you will probably find them in your PHYZZX.LOCAL file.
%
% The example below is used at Princeton.
%
%\def\parsefilename{\expandafter\p@rse\the\picfilename.\endp@rse }
%\def\p@rse#1.#2\endp@rse{\if"#2"\expandafter\picfilename
%        \expandafter{\the\picfilename.fig}\fi }
%
%\newread\pictureread
%\def\redopicturebox{\expandafter\openin\expandafter\pictureread
%                    \the\picfilename
%   \ifeof\pictureread \errhelp=\disabledpictures
%     \edef\n@xt{\errmessage={Cannot find file \the\picfilename}\n@xt
%     \let\pictureboxdefinition=\relax \else
%    \read\pictureread to\y@p \read\pictureread to\y@p
%    \read\pictureread to\x@p \read\pictureread to\y@m
%    \read\pictureread to\x@m \closein\pictureread
%    \p@cht=\y@p truein \advance\p@cht by -\y@m truein
%    \advance\yshift by \y@p truein
%    \p@cwd=\x@p truein \advance\p@cwd by -\x@m truein
%    \advance\xshift by \x@m truein
%    \edef\picturedefinition{\special{pos,inc=\the\picfilename}}%
%    \fi }
%
%
%%%%%%%%%%%%%%%%%%%%%%%%%%%%%%%%%%%%%%%%%%%%%%%%%%%%%%%%%%%%%%%%%%%%%%%%
\input epsf
%%%%%%%%%%%%%%%%%%%%%%%%%%
\def\IR{{\hbox{{\rm I}\kern-.2em\hbox{\rm R}}}}
\def\IB{{\hbox{{\rm I}\kern-.2em\hbox{\rm B}}}}
\def\IN{{\hbox{{\rm I}\kern-.2em\hbox{\rm N}}}}
\def\IC{{\ \hbox{{\rm I}\kern-.6em\hbox{\bf C}}}}

\def\IZ{{\hbox{{\rm Z}\kern-.4em\hbox{\rm Z}}}}
\def\to{\rightarrow}
\def\d{{\rm d}}
\def\underarrow#1{\vbox{\ialign{##\crcr$\hfil\displaystyle
{#1}\hfil$\crcr\noalign{\kern1pt
\nointerlineskip}$\longrightarrow$\crcr}}}
% use of underarrow
%A~~~\underarrow{a}~~~B
%
\def\d{{\rm d}}
\input phyzzx
\overfullrule=0pt
\tolerance=5000
\overfullrule=0pt
\twelvepoint

\twelvepoint
\pubnum{IASSNS-HEP-92/45}
\date{July, 1992}
\titlepage
\title{CHERN-SIMONS GAUGE THEORY AS A STRING THEORY}
\vglue-.25in
\author{Edward Witten\foot{Research supported in part by NSF Grant
PHY91-06210.}}
\medskip
\address{School of Natural Sciences
\break Institute for Advanced Study
\break Olden Lane
\break Princeton, NJ 08540}
\bigskip
\abstract{
Certain two dimensional topological field theories can be interpreted as
string theory backgrounds in which the usual decoupling of ghosts and matter
does not hold.  Like ordinary string models, these can sometimes
be given space-time interpretations.  For instance, three-dimensional
Chern-Simons gauge theory can arise as a string theory.  The world-sheet
model in this case involves a topological sigma model.
Instanton contributions to the sigma model give rise to Wilson line insertions
in the space-time Chern-Simons theory.  A certain holomorphic
analog of Chern-Simons theory can also arise as a string theory.}
\endpage
\chapter{Introduction}

\REF\topsig{E. Witten. ``Topological Sigma Models,'' Comm. Math. Phys.
{\bf 118} (1988) 411.}
\REF\gromov{M. Gromov, ``Pseudo-Holomorphic Curves In Symplectic
Manifolds,'' Invent. Math. {\bf 82} (1985) 307.}
\REF\floer{A. Floer, ``Symplectic Fixed Points And Holomorphic Spheres,''
Comm. Math. Phys. {\bf 120} (1989) 575.}
\REF\axelrod{S. Axelrod and I. M. Singer, ``Chern-Simons Perturbation Theory,''
MIT preprint (1991).}
\REF\kontsevich{M. Kontsevich, lecture at Institute for Advanced Study
(January, 1992).}
\REF\mirror{E. Witten, ``Mirror Manifolds and Topological Field Theory,''
in {\it Essays On Mirror Manifolds}, ed. S.-T. Yau (International Press,
1992).}
In this paper, I will describe how Chern-Simons gauge theory in three
dimensions can be viewed as a string theory.  The string theory in question
will be constructed using a topological sigma model [\topsig]
(related to Floer/Gromov theory)
in which the target
space is $T^*M$, $M$ being a three-manifold.
The perturbation theory of this string theory will coincide with
Chern-Simons perturbation theory, in the form that this has been
presented by Axelrod and Singer [\axelrod] and further studied by
Kontsevich [\kontsevich].  Mathematically, the idea is roughly that
there are no instantons with target space $T^*M$ and boundary values in
$M$, so in an appropriate topological field theory,
the usual counting of instantons is vacuous.  However, there are
virtual instantons at infinity; their proper counting leads to Chern-Simons
perturbation theory.  Chern-Simons theory enters in this particular
string theory in much the same way that ordinary space-time physics
(with general relativity as the long wavelength limit) arises in
conventional string theory.

Physically, one might take the following as the starting point.
String theorists usually construct two dimensional field theories describing
particular classical solutions of string theory by constructing
a ``matter'' system, of total central charge 26, and coupling it
to the ghosts, of central charge $-26$.  The vanishing total
central charge
ensures the existence of a BRST operator $Q$, obeying $Q^2=0$,
and playing a crucial role in world-sheet and space-time gauge invariance.
\REF\siegel{W. Siegel, ``Covariantly Second-Quantized Strings, II,III,''
Phys. Lett. {\bf 151B} (1985) 391,396.}
One knows, however [\siegel], that $Q$ can be interpreted as a generator
of linearized gauge transformations, mixing ghosts and matter,
so the assumption that the matter and ghosts are decoupled cannot
be valid as a fundamental principle; it is merely a (partial) gauge condition
and very likely cannot be imposed at all in some situations, perhaps
time dependent ones.

\REF\elitzur{S. Elitzur, A. Forge, and E. Rabinovici, ``On Effective
Theories Of Topological Strings,'' preprint CERN-TH.6326 (1991).}
So in \S2, we will
ask what is left of the standard structure if one does
not assume decoupling of matter from ghosts.
{}From the discussion, it will become obvious that exotic realizations
of the same basic structure can be constructed using topological sigma
models; we therefore do this in \S3.
The resulting models can be considered for either open or closed strings,
but the open string case is in some ways easier to understand.
In \S4, we use open-string field theory as a short-cut to determining
the space-time interpretation of the open-string version
of our models, with the result - alluded to above -
that one type of model is equivalent in perturbation theory to three
dimensional Chern-Simons gauge theory.
This is possibly the first time that the background independent, gauge
invariant
space-time interpretation of a string theory has been completely determined.
(But see [\elitzur] for a previous
investigation of the space-time interpretation of some topological
string theories of a rather different flavor.)
As one application, our
result explains certain observations by Kontsevich about Chern-Simons
gauge theory, and we will make
a small digression on that account.  Another version of the theory
has for its space-time interpretation a sort of holomorphic version
of Chern-Simons theory.
In \S5, we attempt to discuss the closed string sector (which among other
things
should be more closely related to usual manifestations of mirror symmetry).
This concluding section is
brief since I do not understand it.

To keep this paper within reasonable length, it has not been possible
to give a self-contained explanation of all of the relevant background.
The relevant material on topological sigma models can
be found in [\mirror]; for their coupling to topological gravity see
[\topsig].  The relation of Riemann surfaces to
gauge theories will be briefly recalled
presently.
Apart from this, it is helpful to have some familiarity with construction
of string models, the relation between world-sheet and space-time
physics,  string field theory, and Chern-Simons perturbation theory.
The main point of the paper is the unexpected relation of those latter
topics to each other and to topological sigma models.
%%%%%%%%%%%%%%%%%%%% Sample from Klebanov paper
%\epsfxsize=4in
%\topinsert\centerline{\epsfbox{fig1.ps}}
%{\narrower\smallskip\singlespace
%\noindent Fig. 1.
%A small section of triangulated surface. Solid lines denote the triangular
%lattice $\Lambda$, and dotted lines -- the dual lattice $\tilde\Lambda$.
%\smallskip} \endinsert
%\noindent
%%%%%%%%%%%%%%%%%%%%%%%%%%%%%%%%%%%
%%First Figure insert  csg1Xa-b.eps
\epsfxsize=4in
\topinsert\centerline{\epsfbox{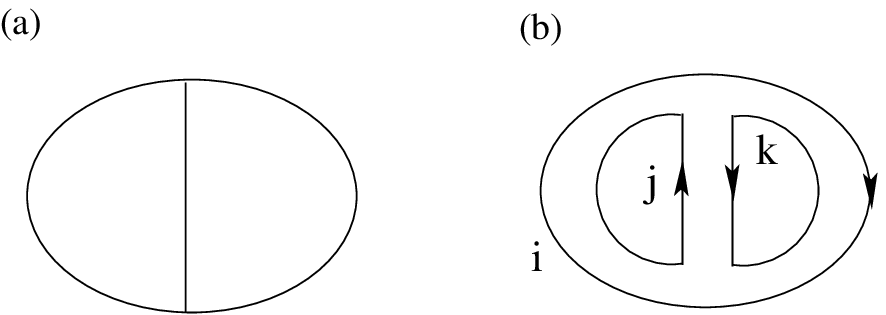}}
{\narrower\smallskip\singlespace
\noindent Fig. 1.
A two loop Feynman diagram (a) and its thickening (b),
in which the boundary components are labeled by gauge indices.
\smallskip} \endinsert
\noindent
\FIG\lto{}
%\FIG\lto{A two loop Feynman diagram (a) and its thickening (b),
%in which the boundary components are labeled by gauge indices.}
%\FIG\ulto{
%The cell decomposition of open string moduli space
%depends on building Riemann surfaces by gluing together
%flat strips $S_i$ of width $\pi$ and variable length $T_i$, $0\leq T_i\leq
%\infty$.  The strips
%are glued together in groups of three along the dotted lines;
%their midpoints meet in conical
%singularities of deficit angle $-\pi$ that are marked as solid dots.
%The resemblance of this figure to figure (\lto) is central in this paper.}
\FIG\ulto{}
%%%%%%%%%%%%%%%%%%%%%%%%%%%%%%%%%%%
%% Figure 2
\epsfxsize=1.5in
\topinsert\centerline{\epsfbox{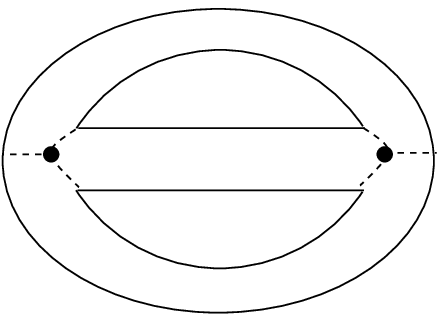}}
{\narrower\smallskip\singlespace
\noindent Fig. 2.
The cell decomposition of open string moduli space
depends on building Riemann surfaces by gluing together
flat strips $S_i$ of width $\pi$ and variable length $T_i$, $0\leq T_i\leq
\infty$.  The strips
are glued together in groups of three along the dotted lines;
their midpoints meet in conical
singularities of deficit angle $-\pi$ that are marked as solid dots.
The resemblance of this figure to figure (\lto) is central in this paper.
\smallskip} \endinsert
\noindent

\subsection{Riemann Surfaces And Gauge Theories}

\REF\thooft{G. 't Hooft, Nucl. Phys. {\bf B72} (1974) 461.}
\REF\wito{E. Witten, ``Two Dimensional Gravity And Intersection Theory
On Moduli Space,'' Surv. Diff. Geom. {\bf 1} (1991) 243.}
Riemann surfaces enter our story in two ways.
On the one hand, when we study two dimensional quantum field theory,
Riemann surfaces are present at the beginning.  Then we wish
to show how gauge theory in the target space emerges.

On the other hand, oriented two dimensional manifolds arise
in gauge theories with $U(N)$ gauge group in the following
fashion, due originally to `t Hooft
[\thooft]; see also [\wito,\S4] for a recent explanation.
($SO(N)$ and $Sp(N)$ can be considered similarly and lead to
not necessarily orientable
Riemann surfaces.)
Though the following general comments also go through for ordinary Yang-Mills
theory in any dimension, let us to be definite consider
the Chern-Simons
action for a $U(N)$-valued connection $A$ on a three manifold $M$:
$$I={k\over 4\pi}\int_M\Tr\left(A\wedge \d A+{2\over 3}A\wedge A\wedge
A\right).
     \eqn\ucov$$
Consider expanding this theory
in perturbation theory in $1/k$, say around the trivial connection.
The successive terms
are all topological invariants.  The coefficient
of $1/k^r$ comes from Feynman diagrams with $r$ loops, and is a function of
$N$.
For given $r$, by considering the dependence on $N$, how many invariants
can be extracted?  This question can be conveniently answered as follows.
The gauge field $A$
is essentially a one-form valued in hermitian $N\times N$ matrices.
Write $A$ as $A^i{}_j$, $i,j=1\dots N$, making the matrix indices explicit
while leaving implicit the fact that $A$ is a one-form.  The interaction
$$\Tr A\wedge A\wedge A = A^i{}_j\wedge A^j{}_k\wedge A^k{}_i\eqn\uvv$$
involves
sewing the ``lower'' index of one $A$ field to the ``upper'' index of the
next.  To exhibit this in drawing Feynman diagrams,
represent the $A$ propagator
not as a line but as a slightly thickened band in which one edge represents
the ``upper'' index and the other line represents the ``lower'' index.
The index flow at the cubic vertex \uvv\ is then neatly incorporated
by smoothly joining the bands, as in figure (\lto(b)).
In the process, the Feynman diagram of figure (\lto(a)) has been replaced
by the Riemann surface $\Sigma$
of figure (\lto(b)).  Each boundary component
is labeled by a gauge index $i$ running from 1 to $N$.  If there are $h$
boundary components, the sum over gauge indices gives a factor of $N^h$.
The coefficients of $N^h/k^r$
are the three-manifold invariants that can be extracted from
Chern-Simons theory with gauge group $U(N)$.

\REF\penner{R. Penner, ``The Teichmuller Space Of A Punctured Surface,''
Comm. Math. Phys. (1987), ``Perturbative Series And The Moduli Space
Of Riemann Surfaces,'' J. Diff. Geom. {\bf 27} (1988) 35.}
\REF\harer{J. Harer, ``The Virtual Cohomological Dimension Of The Mapping
Class Group Of Orientable Surfaces,'' Inv. Math. {\bf 84} (1986) 157.}
\REF\other{B. H. Bowditch and D. B. A. Epstein, ``Natural Triangulations
Associated To A Surface,'' Topology {\bf 27} (1988) 91.}
\REF\ewitten{E. Witten, ``Non-Commutative Geometry And String Field
Theory,'' Nucl. Phys. {\B268} (1986) 253.}
\REF\martinec{S. Giddings, E. Martinec, and E. Witten, ``Modular Invariance
In String Field Theory,'' Phys. Lett. {\bf 176B} (f1986) 362.}
The genus of $\Sigma$ is $g=(r-h+1)/2$.  As $r$ and $h$ vary, $g$ and $h$
vary independently, and $\Sigma$
varies over all topological types of oriented two dimensional surface with
boundary.
Thus, $U(N)$ gauge theory gives us a three manifold invariant $\Gamma_{g,h}(M)$
for every topological type of oriented two dimensional surface with boundary.
On the other hand, in \S3-4 we will consider a two dimensional topological
field theory
(closely related to the counting of almost holomorphic curves) in which
one can conveniently take the target space to be $T^*M$, the cotangent
bundle of a three-manifold $M$.
The partition function $Z_{g,h}(M)$ of this theory
formulated on a Riemann surface of genus $g$ with $h$ holes is, again,
a three-manifold invariant depending on $g$ and $h$.  Our main conclusion
is that $\Gamma_{g,h}(M)=Z_{g,h}(M)$.  In reaching this conclusion,
the link between the Riemann surfaces of the two dimensional field
theory and those of $U(N)$ gauge theory is provided by the fatgraph
[\penner--\other]
or open string field theory [\ewitten,\martinec]
description of the moduli space of Riemann
surfaces with boundary; in this description, complex Riemann surfaces
are built up from pictures similar to that of figure (\lto(b)),
as we will recall in more detail in \S4.

\chapter{Axioms}

In this section, we will discuss what remains of the usual structures
in string theory if one does not assume that ghosts and matter are
decoupled.
$\Sigma$  will be a Riemann surface of genus $g$
with local coordinates $x^\alpha,\,\,\,\alpha=1,2$.
The symbols $J$ and $h$ will be used to denote a complex structure and
a metric on $\Sigma$; of course a metric determines a complex structure.
The space of all complex structures will be called ${\cal J}$; the group
of diffeomorphisms of $\Sigma$ will be called ${\cal G}$.  The quotient
${\cal J}/{\cal G}$ is
the moduli space of complex structures on $\Sigma$.
The space of metrics on $\Sigma$  will be called ${\cal K}$.

Usually one considers conformally invariant world-sheeet theories
consisting
of ``matter,'' of central charge $c=26$, and ``ghosts,''
of $c=-26$.  As explained in the introduction, we want to drop the
assumption of this decoupling; so if conformal invariance is retained,
we preserve the fact that the total
central charge is $c=0$, but no longer build this from a cancellation between
different contributions.
We will, however, not necessarily retain the assumption of conformal
invariance.

\def\A{{\bf A}}
\def\B{{\bf B}}
One part of the structure that must be kept for all that follows
is the existence of a BRST operator $Q$ with $Q^2=0$, and with the
further property that the stress-tensor $T_{\alpha\beta}$ can be written
as
$$T_{\alpha\beta}=\{Q,b_{\alpha\beta}\}, \eqn\totallo$$
for some field $b_{\alpha\beta}$ -- which in the usual case is called
the antighost field.  This ensures general covariance -- or topological
invariance -- of the two dimensional theory.  In the conformally invariant
case, $b$ and $T$ are traceless.  The stress tensor is conserved,
$D_\alpha T^{\alpha}{}_\beta=0$, and so we will assume that the $b$ field
obeys
$$ D_\alpha b^{\alpha}{}_\beta=0 \eqn\otallo$$
which when $b$ is traceless is its usual equation of motion.
In addition, in the usual constructions,
there are no short-distance singularities in operator products of
$b$ fields, and there are only the standard short-distance singularities
in products of $b$ and $T$.
Concerning the generalization, topological sigma models are an example
showing that these assumptions are too strong.  In those models,
there is a delta
function contact term in the $b\cdot T$ product.  (This results
in the $\rho\rho\psi\psi$ term in the coupling to topological gravity,
eqn. (4.18) of [\topsig].)
I do not know what sort of singularities should be allowed in $b\cdot b$
and $b\cdot T$ operator products, in general, except to say that they
must be such that a coupling of the model to topological gravity must
be possible.

Another feature of the usual case that we wish to preserve
is that there is a ``ghost
number'' operator $G$ (mathematically it would usually be called the
dimension),
with
$Q$ and $b$ having $G=1$ and $G=-1$, respectively.
Moreover, we wish to preserve the usual fact that the ``ghost number of
the vacuum'' is $-3\chi(\Sigma)$ where $\chi(\Sigma)$ is the Euler
characteristic of $\Sigma$; this means that only an
operator product of total ghost number
$3\chi(\Sigma)$ can have a non-zero
vacuum expectation value.
For instance, for an orientable Riemann surface of genus $g$, the case
that we will consider in this section for definiteness,
$3\chi(\Sigma)=-(6g-6)$.

A particularly
important case (in genus $\geq 2$ for simplicity, to avoid
ghost zero modes) is the expectation value of a product of $6g-6$
$b$'s.  Let $\delta^{(k)}h$, $k=1,\dots ,6g-6$ be $6g-6$ variations of
$h$ (that is, $6g-6$ tangent vectors to the space ${\cal K}$ of metrics).
Let
$$b^{(k)}=\int_\Sigma \d^2x\sqrt h h^{\alpha\alpha'}h^{\beta\beta'}
\delta^{(k)}h_{\alpha\beta}\cdot b_{\alpha'\beta'}\eqn\gogo$$
be the corresponding modes of $b$, and let
$$\Theta(\delta^{(1)}h,\dots,\delta^{(6g-6)}h)=\langle b^{(1)}\dots
b^{(6g-6)}\rangle ,       \eqn\cocn$$
with $\langle~~~~\rangle$ the expectation value in the ``measure'' determined
by the genus $g$ world-sheet path integral.
By fermi statistics, $\Theta$ is skew-symmetric in the $\delta^{(k)}h$'s
and so can be interpreted as a $6g-6$ form on ${\cal K}$.
In the usual case, one proves that $\Theta$ is closed starting with
$$0=\langle\{Q,b^{(1)}\dots b^{(6g-5)}\}\rangle
=\sum_{j=1}^{6g-5}(-1)^{j-1}\langle b^{(1)} \dots \{Q,b^{(j)}\}\dots
b^{(6g-5)}\rangle. \eqn\noffo$$
Then using \totallo\ to write $\{Q,b^{(j)}\}$ as a moment of $T$,
and interpreting the insertion of $T$ as a derivative on ${\cal K}$,
one can interpret \noffo\ as a first order differential equation for $\Theta$
which amounts to
$$ \d\Theta=0. \eqn\goffo$$
This proof goes through in general if the $b\cdot b$ and $b\cdot T$
operator products are standard; however, topological sigma models are an
example
in which the definition of $\Theta$ must be modified by addition of contact
terms to ensure $\d\Theta = 0$.  A successful coupling of the model to
topological gravity will always ensure the existence of a suitable modification
(by addition of contact terms) of the definition of $\Theta$.

$\Theta$ is obviously diffeomorphism invariant.  Moreover, in the usual
case it is ``basic'';
that is, it vanishes if any $\delta^{(k)}h$ is of the form
$\delta^{(k)}h_{\alpha\beta}= D_\alpha v_\beta+D_\beta v_\alpha$ induced by
an infinitesimal diffeomorphism $x^\alpha\to x^\alpha+\epsilon v^\alpha$.
This is proved by integrating by parts, using \otallo\ and the absence
of short distance singularities of the $b$'s.
In general (because of the relation of the topological gravity multiplet
to the equivariant cohomology of the diffeomorphism group), a successful
coupling to topological gravity will always ensure that $\Theta$ is basic.

The fact that $\Theta$ is basic and diffeomorphism invariant means that
it can be interpreted as the pullback of a closed $6g-6$ form $\nu$
on the space ${\cal K}/{\cal G}$ of metrics modulo diffeomorphisms.
In the conformally invariant case, $\Theta$ is actually a pullback from
a form $\mu$ on
the moduli space ${\cal M}$ of complex structures; as
${\cal M}$ is $6g-6$ dimensional,
$\mu$ is a top form or measure on ${\cal M}$.
If conformal invariance does not hold, there is no natural way
to construct from $\mu$ a measure on ${\cal M}$, but the fact that
$\nu$ is closed, $\d\nu=0$, is almost as good.  It means that
if $s$ is a section of the bundle ${\cal K}/{\cal G}\rightarrow {\cal M}$,
and $\mu=s^*(\nu)$, then $\mu$ is independent of the choice of $s$, modulo
an exact form.  $\mu$ is the desired $6g-6$-form on ${\cal M}$.

\subsection{Vertex Operators And Ghosts}

\REF\nelson{J. Distler and P. Nelson, ``Topological Couplings And Contact
Terms In 2d Field Theory,'' Comm. Math. Phys. {\bf 138} (1991) 273.}
The most traditional
string theory vertex operators can be written $W=c\overline c V$
where $c$ and $\overline c$ are the ghosts and $V$ is a spin $(1,1)$
conformal field constructed from matter fields only.  It is well
known in the operator formalism [\nelson] that these conditions can
be relaxed.  The only really essential properties of $W$ in conventional
string theory are that it have ghost number 2, spin zero, and be annihilated
by $Q$, and that the Fock space state corresponding to $W$ should be
annihilated
by an operator usually written as $b_0{}^-=b_0-\overline b_0$
(here $b_0$ and $\overline b_0$ are the zero modes of left- and right-moving
ghosts).  The requirement that $W$ have spin zero and
be annihilated by $b_0{}^-$ ensures that the analogs of $\Theta$
defined with an insertion of $W$ have the essential properties that entered
above  (diffeomorphism invariant and basic).
The ghost number two condition shifts the ghost number of
the vacuum, leading in a natural way to measures on the moduli space of
Riemann surfaces with marked points.

These conditions make sense in our abstract setting, and can be borrowed
bodily, except that if we do not assume conformal invariance, the definition
of $b_0{}^-$ must be written $b_0{}^-=\oint \d x^\alpha b_{\alpha 0}$.
(The integration is over a parametrized circle and ``0'' is the normal
direction to the circle.)
Note, therefore, that in including vertex operators, we need make
no explicit mention of the ghosts; only the product $W= c\overline c V$,
and not the separate factors in that product, needs to be generalized.
Where, therefore, do the ghosts appear in the discussion?
Of course, the antighosts entered in the definition of the measure for
$g\geq 2$; but what about the ghosts?

We perhaps could ``find'' the ghosts by trying to generalize the usual
definition of the genus one measure -- as $c$ and $\overline c$ each
has a zero mode on a surface of genus one.  To get a different perspective,
I will instead discuss the issue from the standpoint of closed string
field theory.

\REF\zwiebach{B. Zwiebach, ``Closed String Field Theory: Quantum
Action And The BV Master Equation,'' preprint IASSNS-HEP-92/41 (June, 1992).}
In closed string field theory as we know it today (see [\zwiebach] for a
review),
the string field $\Psi$
is a vector in a string Hilbert space obtained by quantization on a circle
$S$.  Picking a parametrization of the circle, let $L_0{}^-=L_0-\overline
L_0$ be the generator of a rotation of the circle, and let
$c_0{}^-=(c_0-\overline c_0)/2$.  $\Psi$ is required to obey
$$L_0{}^-\Psi
=b_0{}^-\Psi = 0.\eqn\hodo$$
The quadratic part of the Lagrangian is
$$\left(\Psi,c_0{}^-Q\Psi\right), \eqn\humbo$$
and the linearized gauge invariance is
$$\delta\Psi=Q\epsilon, \eqn\jumbo$$
where $\epsilon$ has ghost number 1 and
$$b_0{}^-\epsilon = 0. \eqn\cumbo$$
This is proved to be a gauge invariance as follows.
In the usual theory, \cumbo\ implies that
$$\epsilon=b_0{}^-\alpha \eqn\umbo$$
for some $\alpha$.  Inserting this in
\humbo, and using
$$\{b_0{}^-,c_0{}^-\} = 1, \eqn\mbo$$
together with
$$ \{Q,b_0{}^-\}=L_0{}^-\eqn\nbo$$
(which is a consequence of \totallo) along with
\hodo, one verifies the invariance
of \humbo.

What are the essential points that should be retained if the ghosts
and matter fields are not decoupled?  In conventional string theory,
one has
$$ \left(b_0{}^-\right)^2 = 0, \eqn\vxo$$
a consequence of the absence of singularities in the $b\cdot b$ operator
products.  This property also holds in topological sigma models, and it
seems reasonable to insist on it in general even if one permits some
kind of $b\cdot b$ singularities.
The ability to write $\epsilon$ as in \umbo\ is then the assertion that
the cohomology of $b_0{}^-$ vanishes in ghost number 1.  If the cohomology
of $b_0{}^-$ is altogether zero, then there exists an operator
$c_0{}^-$ obeying \mbo.  Moreover, $c_0{}^-$ is uniquely determined
modulo $c_0{}^-\rightarrow c_0{}^-+\{b_0{}^-,f\}$ for some $f$;
using \hodo\ and \nbo, one sees that the Lagrangian
\humbo\ is invariant under such a shift of $c_0{}^-$.
Thus, \vxo\ and the trivial cohomology of $b_0{}^-$ are sufficient
requirements for free closed string field theory (and I believe also
for the interacting case).

What if the cohomology of $b_0{}^-$ is not zero?  Then using \nbo\
and the fact that the
cohomology can be represented by states invariant under the compact
group generated by $L_0{}^-$ (and in any case since the physical
field $\Psi$ is required to have this invariance),
we see that $Q$ generates a linear transformation of the cohomology of
$b_0{}^-$.  If this linear transformation is zero, there will be an operator
$c_0{}^-$ obeying not \mbo\ but
$$ \{b_0{}^-,c_0{}^-\}Q=Q\{b_0{}^-,c_0{}^-\}= Q . \eqn\ofto$$
This actually is enough to ensure the gauge invariance of
the free Lagrangian \humbo.
(If the cohomology of $b_0{}^-$ is
nonzero in ghost number one,
one cannot use \umbo; but the fact that $Q$ annihilates the cohomology
of $b_0{}^-$ means that for $\epsilon$ annihilated by $b_0{}^-$ and $L_0{}^-$,
one can write $Q\epsilon=b_0{}^-\beta$ for some $\beta$; this is good enough.)
Topological sigma models give an example in which the cohomology of $b_0{}^-$
is non-zero, but annihilated by $Q$.
We therefore can write down closed string field theory in this case,
but as we will
see in \S5, it seems difficult to understand it.

\subsection{A Conundrum}

The conundrum that I want to state is obvious.  The properties of
conventional string backgrounds that I have cited are the only ones
I know  of that make sense in the general case
in which ghosts and matter are not decoupled.  Yet they are so general
as to permit bizarre realizations, like the topological sigma models
that we will consider presently.  Are they adequate and if not how should
they be supplemented?

\REF\dixon{L. Dixon, lecture at Princeton University, ca. 1987.}
The following example will perhaps serve to sharpen the puzzle.
In Type II superstring theory, it is usually supposed that a classical
solution is described by a superconformal world-sheet theory.  One therefore
might expect and hope to supplement the above discussion with general
properties of superconformal symmetry.  Yet it has been pointed
out [\dixon] that Ramond-Ramond vertex operators do not commute
with the left- or right-moving superconformal currents, and consequently
that when such vertex operators are added to the world-sheet
Lagrangian (as one would expect in a generic time-dependent situation),
the world-sheet theory is not superconformal.  Therefore, the
structure of Type II backgrounds, at the general level of our above discussion,
looks hard to distinguish from that of bosonic string theory.
If we do not wish to claim that bosonic strings and Type II superstrings
are the same theory (and I would be skeptical of that interpretation),
we apparently must find general properties of the world-sheet theory
that go well beyond the ones I have cited.

\chapter{Realizations Via Topological Field Theories}

\REF\vafa{C. Vafa, ``Topological Mirrors And Quantum Rings,'' in
{\it Essays on Mirror Manifolds}, ed  S.-T. Yau (International Press, 1992).}
\REF\labastida{J. M. F. Labastida and P. M. Llatas, ``Topological
Matter in Two Dimensions'' (preprint, 1991).}
We will now use topological sigma models to make
some realizations of the structure just explained.
There are two classes of such models [\mirror,\vafa,\labastida],
which I will call the $\A$ and $\B$ models following some of the papers
just cited.

These models govern maps from a Riemann surface $\Sigma$ to a target space
$X$, which must be presented with an
almost complex structure, in the case of the $\A$
model, or an actual complex structure, for the $\B$ model.
The $\A$ model is concerned with almost holomorphic maps from
$\Sigma$ to $X$, while the $\B$ model is related to periods of differential
forms on $X$.
The $\B$ model is only well-defined for $X$ obeying the Calabi-Yau
condition $c_1(X)=0$.
The $\A$ model is defined without that condition, but obeys the axioms
of \S2 (definite ghost number
of the vacuum) only if $c_1(X)=0$.  Therefore, we will limit ourselves
to Calabi-Yau target spaces.  This is, of course, also the case in which
the sigma model is conformally invariant.

For either the $\A$ or $\B$ models, the ghost number of the vacuum is
$-d\cdot\chi(\Sigma)$, with $d=\dim_\IC(X)$.  This coincides with the
desired valued $-3\chi(\Sigma)$
that played such an important role in \S2 precisely if
$d=3$, and therefore we will limit ourselves to this case.  It is
curious to note that this is the same value of $d$ that arises in the usual
``physical'' applications of Calabi-Yau manifolds in superstring
compactification.

General tangent space indices to $X$ will be written as $I,J,K$, while indices
of type $(1,0)$ or $(0,1)$ will be written as $i,j,k$ or $\overline i,
\overline j, \overline k$, respectively.  We consider $X$ as a Kahler manifold
endowed with a Ricci-flat Kahler metric $g_{i \bar j}$.
The bosonic fields of the $\A$ or $\B$ models are simply a map
$\Phi:\Sigma\to X$; if we pick local coordinates $\phi^I$ on $X$, this
map can be described by giving functions $\phi^I(x^\alpha)$.
It is also convenient to pick a local complex coordinate $z$ on $\Sigma$.

\subsection{The $\A$ Model}

The detailed construction of the $\A$ and $\B$ models is explained in
[\mirror].
In the $\A$ model, the fermi fields are a section $\chi^I$ of $\Phi^*(TX)$,
and a one-form $\psi$ with values in $\Phi^*(TX)$. $\psi$
obeys a self-duality
condition which says that its
$(1,0)$ part $\psi_z^{\overline i}$ has values in $\Phi^*(T^{0,1}X)$,
and its $(0,1)$ form $\psi_{\overline z}^i$ has values in $\Phi^*(T^{1,0}X)$.
The BRST transformation laws are
$$\eqalign{
\delta\phi^I  &   = i\alpha \chi^I \cr
\delta\chi^I&  = 0 \cr
\delta\psi_z^{\bar i}& = -\alpha\partial_z\phi^{\bar i}-
i\alpha\chi^{\bar j}\Gamma^{\bar i}_{\bar j\overline m}
         \psi_z^{\overline  m} \cr
\delta\psi_{\overline z}^{i} & = -\alpha\partial_{\overline z}
\phi^{i}-i\alpha\chi^{j}\Gamma^{i}
_{jm}\psi_{\overline z}^m\cr} \eqn\gulfo$$
with $\alpha$ an anticommuting parameter and $\Gamma$ the affine connnection
of $X$.  The BRST operator $Q$ is defined
by writing $\delta \Lambda = -i\alpha\{Q,\Lambda\}$ for any field $\Lambda$.
The Lagrangian can be written in the form $L=i\{Q,V\}$ with any $V$ such
that $L$ is nondegenerate.  A suitable choice of $V$
is
$$V=t\int_\Sigma \d^2z\,\,\,\,
g_{\overline ij}\left(\psi_z ^{\overline i}\partial
_{\bar z}\phi^j+\partial_z\phi^{\overline i}\psi_{\overline z}^j\right)
\eqn\forlateruse$$
(with $\d^2z =|\d z\wedge \d\overline z|$) and gives
$$L=2t\int_\Sigma \d^2z \left({1\over 2}g_{IJ}\partial_z\phi^I
\partial_{\overline z}
\phi^J+i\psi_{z}^{\overline i}D_{\overline z}\chi^ig_{\overline i i}
+i\psi_{\overline z}^{i}D_{z}\chi^{\overline i}g_{\overline i i}
-R_{i\overline i j\overline j}\psi_{\overline z}^i\psi_z^{\overline i}
\chi^j\chi^{\overline j}\right) \eqn\ddalfo$$
with $t$ a constant.

The fact that $L=i\{Q,V\}$ means that the $t$ dependence
(and dependence on the target space metric)
of \ddalfo\
is of the form $\{Q,\dots\}$ and so does not affect the BRST
invariant physics.
As weak coupling means large $t$, the fact that there is no $t$ dependence
means that weak coupling is exact.

The dependence on the
metric of $\Sigma$ is likewise of the form $\{Q,\dots\}$, so
we can
write $T_{\alpha\beta}=\{Q,b_{\alpha\beta}\}$
where at the classical level
$$ b_{\alpha\beta}=itg_{IJ}\left(\psi_\alpha^I\partial_\beta \phi^J
+\psi_\beta^I\partial_\alpha\phi^J-h_{\alpha\beta}h^{\sigma\tau}
\psi_\sigma^I\partial_\tau\phi^J\right).   \eqn\ormo$$
Quantum mechanically, the formula for $b$ requires modification except in the
Calabi-Yau case ($T$ and $b$ will no longer be traceless), but an appropriate
$b$ still exists, as the theory admits a higher derivative regularization
preserving the fermionic symmetry.

\subsection{The $\B$ Model}

In the $\B$ model, the fermi fields include first of all
sections $\eta^{\overline i}$,
$\theta^{\overline i}$ of $\Phi^*(T^{0,1}X)$; actually, it is convenient
to write the formulas in terms of not $\theta^{\bar i}$ but
$\theta_j=g_{j\overline i}\theta^{\overline i}$.
The other fermi fields are a one-form $\rho^i$ with values in
$\Phi^*(T^{1,0}X)$.
The transformation laws are
$$\eqalign{\delta \phi^i    &   =      0 \cr
           \delta \phi^{\overline i} & =i\alpha\eta^{\bar i} \cr
            \delta \eta^{\bar i} & =\delta \theta_i = 0 \cr
            \delta \rho^i & = -\alpha \,\,\d\phi^i  \cr} \eqn\newtr$$
and the Lagrangian is again of the form $L=i\{Q,V\}$; for a suitable
$V$, one gets
$$\eqalign{
L=&t\int_\Sigma d^2z\left(g_{IJ}\partial_z\phi^I\partial_{\bar z}\phi^J
+i\eta^{\bar i}(D_z\rho_{\bar z}^i+D_{\bar z}\rho_z^i  )g_{i\bar i}
\right.\cr &\left.
\qquad +i\theta_i(D_{\bar z}\rho_z{}^i-D_z\rho_{\bar z}{}^i)
+R_{i\bar ij\bar j}\rho_z^i\rho_{\bar z}^j\eta^{\bar i}\theta_k g^{k\bar j}
\right).\cr} \eqn\xolfo$$
The physics is independent of the coupling parameter $t$ for the same
reason as in the $\A$ model.

In this case, we can write $T_{\alpha\beta}=\{Q,b_{\alpha\beta}\}$ with
$$ b_{\alpha\beta}=it g_{IJ}\left(\rho_\alpha^I\partial_\beta \phi^J
+\rho_\beta^I\partial_\alpha\phi^J-h_{\alpha\beta}h^{\sigma\tau}
\rho_\sigma^I\partial_\tau\phi^J\right).   \eqn\ormo$$

\section{Boundary Conditions}

\REF\horava{P. Horava, ``Equivariant Topological Sigma Models,'' preprint
(1991).}
In our applications, we will want to formulate these theories on
Riemann surfaces $\Sigma$ that may have non-empty boundary.
In doing
so, we require a boundary condition.
\foot{For some generalities about topological field theories on Riemann
surfaces
with boundary, and some specifics about the $\A$ model, see [\horava].}
We want local, elliptic boundary conditions that preserve the fermionic
symmetry.  I will explain natural boundary conditions for both the
$\A$ and $\B$ model, but there may very well be other natural boundary
conditions -- perhaps required for mirror symmetry -- that should be
considered.

The $\A$ model is closely related
to Floer theory, and the boundary condition that I want to consider
for the $\A$ model is the one that was studied by Floer.
\foot{Readers to whom these boundary conditions may seem strange at first
sight are invited to try to find other local, elliptic, $Q$-invariant boundary
conditions for the $\A$ model.}
For each component $C_i$ of $\partial \Sigma$, we pick a Lagrangian
submanifold $M_i$ of the Kahler manifold $X$.  For instance, in the nicest
case, there may be a real involution $\tau_i$ of $X$, and $M_i$ may
be a component of the fixed point set of $\tau_i$.
Let $TM_i$ and $NM_i$ be the tangent and normal bundles to $M_i$ in
$X$; we regard them as the real and imaginary subbundles of $TX|_{M_i}$.
Then regarding
$C_i$ as a real locus in the complex Riemann surface $\Sigma$, we require
that the boundary values should be real in the following sense:
$\Phi|_{C_i}$ is real, that is it maps $C_i$ to $M_i$; the normal
derivative to $\Phi$ at $C_i$ is imaginary (it takes values in $\Phi^*(NM_i)$);
$\chi$ and the pullback of $\psi$ to $M_i$ are real (they take values in
$\Phi^*(TM_i)$).
These boundary conditions make sense even
if $M_i$ does not arise as the fixed point set of  a real involution.

For the $\B$ model, we pick instead ``free'' boundary conditions
that do not require anything analogous to the choice of the $M_i$.
We require that the normal derivative of $\Phi$ vanishes on $\partial\Sigma$,
that $\theta$ vanishes on $\partial \Sigma$, and that the pullback to
$\partial\Sigma$ of
$\star \rho$ vanishes ($\star$ being here the Hodge star operator).

\section{Boundary Terms}

Now, we further wish to couple this system to target space gauge fields
in the sense of string theory, that is to gauge fields on $X$.
This will be done by introducing Chan-Paton factors, that is by coupling
the gauge fields to ``charges'' that propagate on $\partial\Sigma$.
These charges are string theory analogs of the labels on the boundaries
in figure (\lto(b)), and will reduce to the latter when we make contact
with target space Chern-Simons theory.
Just as in the construction leading to figure (\lto(b)), we must take
the gauge group to be $U(N)$ if we want to consider oriented Riemann surfaces
only; $SO(N)$ and $Sp(N)$ are possible if one wishes to permit unoriented
surfaces.  The restriction on the gauge group is not needed in the
preliminary discussion of the present section but is essential in \S4.

The quantum version of either
of the models we have introduced
is described by a Feynman path integral
$$\int D\Psi_i\,\,\,\,\exp(-L(\Psi_i)), \eqn\xurgo$$
where $\Psi_i$ are the various fields and $L$ is the Lagrangian.
Let now $A=A_I\d\phi^I$ be a connection with structure group $U(N)$
on a rank $N$ complex vector bundle $E$ over $X$.  Let $\Sigma $ be an
oriented Riemann surface whose boundary is a disjoint union of circles
$C_i$.  The orientation of $\Sigma$ induces an orientation of each $C_i$.
Given $\Phi:\Sigma\to X$,
for each $C_i$ we can take the trace of the holonomy of $\Phi^*(A)$
around $C_i$; we write this as
$$\Tr P\exp\oint_{C_i}\Phi^*(A). \eqn\bbc$$
(The trace is taken in the defining $N$ dimensional representation of $U(N)$.)
Then, tentatively, the coupling to gauge fields in the target space is
accomplished by replacing \xurgo\ by
$$\int D\Psi_i\,\,\,\,\exp(-L(\Psi_i))\cdot\prod_i\Tr P\exp\oint_{C_i}
\Phi^*(A).
 \eqn\xpurgo$$
(For the $\B$ model, this will require some modification.)
If $A$ is trivial, this is simply a factor of $N$ for each boundary
component, just as in the evaluation of the Feynman amplitude of figure
(\lto(b)).

We must determine whether this modification of the path integral
preserves the fermionic symmetry.
In general the variation of the trace of the holonomy about a circle
$C$ is
$$\delta  \Tr P
\exp\oint_C\Phi^*(A)=\Tr \oint_C \delta \phi^I {\d\phi^J\over
\d\tau}F_{IJ}(\tau)\d\tau \cdot P\exp\oint_{C;\tau} \Phi^*(A) .\eqn\hobo$$
Here $\tau$ is a coordinate on $C$, and $\exp\oint_{C;\tau}\Phi^*(A)$ is
the holonomy of $\Phi^*(A)$ around $C$, starting and ending at $\tau$;
and $F_{IJ}$ (or more fastidiously $\Phi^*(F_{IJ})$) are the components
of the pullback by $\Phi$ of the space-time curvature $F=\d A+A\wedge A$.

\subsection{The $\A$ Model}

In the case of the $\A$ model, since $\Phi^*(C_i)\subset M_i$, the
bundle $E$ and connection $A$ actually
need only be defined on the union of the
$M_i$, not on all of $X$.

Inserting the $\A$ model transformation law
$\delta\phi^I=i\alpha\chi^I$ in \hobo, we see that
the holonomy factors in \xpurgo\ are invariant under the fermionic symmetry
if and only if the space-time curvature
$F$ vanishes.  Thus, in the $\A$ model, it is possible
to couple only to {\it flat} connections on the $M_i$.
We will give this a more intuitive explanation in \S4, where we will
see that the target space physics of the $\A$ model is Chern-Simons
gauge theory on the $M_i$.
The classical solutions of Chern-Simons gauge theory
are precisely the flat connections.  Thus, the requirement
of the $\A $ model that the target space connection
must be flat is a special case
of the fact that in string theory, the background fields that can be
incorporated in the world-sheet theory are always classical
solutions of the space-time theory.

Since $A$ is flat, its role is from some points of view almost trivial.
The homotopy factors in \xpurgo\ depend only on the topology of $\Sigma$
and the choice of a homotopy class of maps $\Phi:\Sigma\to X$, in which
one is evaluating the functional integral.  But deeper aspects of the
theory involve summing over the topology of $\Sigma$ and the homotopy type
of $\Phi$, and then the factors coming from holonomies of a flat connection
on $X$ fit together coherently (to give, as we will see, Chern-Simons
theory expanded around the given flat connection $A$).

\subsection{The $\B$ Model}

The situation in the $\B$ model is similar but more subtle.
Plugging the transformation laws $\delta\phi^i=0$,
$\delta\phi^{\overline i}=i\alpha\eta^{\overline i}$ into \hobo,
we find that the holonomy is invariant under the fermionic symmetry
if and only if the $(1,1)$ and $(0,2)$ parts of the curvature vanish.
The vanishing $(0,2)$ curvature asserts that the $(0,1)$ part of $A$
defines a holomorphic structure on the bundle $E$;
the additional vanishing of the $(1,1)$ curvature
says that $A$ is a holomorphic connection
(locally  it can be represented by a holomorphic one-form of type
$(1,0)$).  The latter condition is rather restrictive and we wish
to eliminate it.
This can be done as follows.
Replace $\Phi^*(A)$ by the following ``improved'' connection on the bundle
$\Phi^*(E)$ over $\Sigma$:
$$ \widetilde A=\Phi^*(A)-i\eta^{\overline i}F_{\overline i j}\rho^j
.\eqn\cocno$$
Then using the transformation laws of the $\B$ model, one readily sees
that for any circle $C\subset \Sigma$, the trace of the holonomy
$$\Tr P\exp\oint_C \widetilde A \eqn\ocno$$
of $\widetilde A$ is invariant under the fermionic symmetry, provided
that the $(0,2)$ part of the curvature of $A$ vanishes, that is provided
$A$ determines a holomorphic structure on $E$.
Thus, the formula \xpurgo\ for coupling external gauge fields to the $\A$
model should be modified, in the case of the $\B$ model, to
$$\int D\Psi_i\,\,\,\,\exp(-L(\Psi_i))\cdot\prod_i\Tr P\exp\oint_{C_i}
\widetilde A.
 \eqn\xxpurgo$$
I will leave it to the interested reader to verify that up to terms of the
form $\{Q,\dots\}$, the coupling to gauge fields in \xxpurgo\ depends only
on the holomorphic structure of $E$ (\ie, the $(0,1)$ part of $A$),
not on the $(1,0)$ part of $A$.
This is in accord with the fact that, as we will see
in \S4, the $\B$ model has for its
classical solutions in space-time the holomorphic vector bundles.
As always, the classical solutions are the objects to which the world-sheet
theories can be coupled.

\section{Large $t$ Limit}

To set up the Hamiltonian version of these theories, for open strings,
we take $\Sigma$ to be an infinite strip $0\leq \sigma\leq \pi$,
$-\infty\leq \tau\leq \infty$ with metric $\d s^2=\d\sigma^2+\d \tau^2$.
We consider the $\A$ and $\B$ models with the boundary conditions
just described.  In the case of the $\A$ model, we use the same Lagrangian
submanifold $M$ at the two ends of the strip.  A quantum Hilbert
space ${\cal H}$
is introduced in the usual way by quantizing on the initial
value surface $\tau=0$.  We want to compute the cohomology of $Q$
and certain aspects of the large
$t$ behavior that will be essential in the next section.

\subsection{The $\A$ Model}

If we write $\psi=\psi_\sigma\d \sigma+\psi_\tau\d\tau$, then, as the
self-duality condition determines $\psi_\sigma$ in terms of $\psi_\tau$,
we can regard $\chi $ and $\psi_\tau$ as the independent fermi variables.
The canonical commutation relations are
$$\eqalign{ \left[{\d\phi^I\over\d\tau}(\sigma),\phi^J(\sigma')\right]
&=-{i\over t}g^{IJ}\delta(\sigma-\sigma')\cr
\{\psi_\tau(\sigma),\chi(\sigma')\}& ={1\over t}\delta(\sigma-\sigma').\cr}
 \eqn\concom$$
The Hilbert space ${\cal H}$ consists of functionals ${\cal A}(\Phi,\dots )$
where now $\Phi$ is a map of the interval $I=[0,\pi]$ to $X$ mapping
$\partial I$ to $M$, and ``$\dots$''
is a subset of half the fermi variables,
depending on a choice of representation of the canonical anticommutators.

The Hamiltonian is
$$L_0=\int_0^\pi\d\sigma \,\,\,T_{00}. \eqn\hamton$$
Using the canonical commutation relations to write $\d\phi/\d\tau$ in terms
of $\delta/\delta\phi$, this can be written
$$L_0={1\over 2}\int_0^\pi\left(-{1\over t}g^{IJ}{\delta^2\over\delta\phi^I
\delta\phi^J(\sigma)}+tg_{IJ}{\d\phi^I\over\d\sigma}{\d\phi^J\over\d\sigma}
\right)+{\rm terms~with~fermions}.\eqn\dufus$$
The fundamental relation $T_{\alpha\beta}=\{Q,b_{\alpha\beta}\}$ implies
that if we introduce the zero mode of the $b$ field,
$$b_0=\int_0^\pi \d\sigma \,\,b_{00}\eqn\amton$$
then
$$L_0=\{Q,b_0\}.     \eqn\gamton$$

\gamton\ implies that
the $Q$ cohomology can be computed in the subspace of ${\cal H}$ annihilated
by $L_0$.  Since the $Q$ cohomology (and everything else of essential interest)
is independent of $t$, it is enough to study the kernel of $L_0$ for large
$t$.  Actually, with other applications in mind, we want to understand
not just the kernel of $L_0$ but all eigenvalues that are of order $1/t$
for large $t$.

Looking at \dufus, we see that such eigenfunctions must be localized near the
region with $\d\phi^I/\d\sigma=0$, that is, the space of constant maps $\Phi
:I\to X$.
Because of the boundary condition that $\partial \Sigma$ is mapped to $M$,
the constant map in question must in fact map $I$ to $M$.
The non-zero modes of the fermions make contributions of order 1 to $L_0$.
Combining these observations, the low-lying eigenfunctions of $L_0$
can be described by a functional ${\cal A}$ of the bose and fermi zero modes,
with other modes in their Fock vacuum.  (One must check that the
energy of this Fock vacuum is zero, but this follows from $Q$-invariance or
better from the supersymmetry of the untwisted theory.)
The bose and fermi zero modes are all tangent to $M$; denote them as
$q^a,\chi^a,
\psi_\tau{}^a, \,a=1\dots 3$, with $q^a$ being coordinates on $M$.
(We reserve the letter $\phi$ for coordinates on $X$.)  The
canonical commutation relations
\concom\ show that $\psi_\tau{}^a$ can be represented as $\partial/\partial
\chi^a$; then ${\cal A}$ reduces to a function ${\cal A}(q^a,\chi^a)$.
This has an expansion in powers of $\chi^a$
$${\cal A}=c(q)+\chi^a A_a(q)+\chi^a\chi^b B_{ab}(q)+
\dots \eqn\omigo$$
in which the successive terms can be interpreted as
$p$-forms on $M$ of degrees
$0\leq p\leq 3$.  If Chan-Paton factors are included in the discussion, these
differential forms take values in ${\rm End}(E)$, the endomorphisms
of some flat vector bundle
$E$ over $M$.

The commutation relations $[Q,\phi^I]=-\chi^I$, $\{Q,\chi^J\}=0$ show that,
if we interpret $\chi^I$ as $-\d\phi^I$, we can identify $Q$ with the exterior
derivative $\d$ on $M$ (with values in ${\rm End}(E)$).  The $Q$ cohomology
is thus $H^*(M,{\rm End}(E))$.

Moreover,
\dufus\ shows that acting on differential forms on $M$, $L_0$ reduces
to
$$L_0={\pi\over 2t}\Delta ,\eqn\ungo$$
with $\Delta=\d\d^*+\d^*\d$ the usual Laplacian on forms. The underlying
relation $L_0=\{Q,b_0\}$ leads one to guess that in the same approximation
$$b_0={\pi\over 2t}\d^*, \eqn\nugo$$
and this can be verified using the commutation relations.

\subsection{The $\B$ Model}

For the $\B$ model we can be brief, as the arguments are so similar.
In the large $t$ limit, the wave-functional ${\cal A}$ reduces to a function
of the zero modes of $\phi^I$ and $\eta^{\overline i}$ (the zero mode
of $\rho$ being represented as $\partial/\partial\eta$).  Expanding
$${\cal A}(\phi^I,\eta^{\overline i})=c(\phi^I)+\eta^{\overline i}A_{\overline
i}(\phi^I)
+\eta^{\overline i}\eta^{\overline j}B_{\overline i\overline j}(\phi^I)
+\dots, \eqn\ucco$$
we see that for the low-lying states, ${\cal A}$ reduces to a sum
of $(0,q)$ forms on $X$ (valued in ${\rm End}(E)$, $E$ being a holomorphic
vector bundle on $X$), for $0\leq q\leq 3$.
The commutation relations $[Q,\phi^i]=0$, $[Q,\phi^{\overline i}]=-\eta^{
\overline i}$, $\{Q,\eta^{\overline i}\}=0$ show that, if we interpret
$\eta^{\overline i}$ as $-\d \phi^{\overline i}$, we can identify $Q$
with the $\overline\partial $ operator (with values in ${\rm End}(E)$).
So the cohomology of $Q$
is $H^{0,*}(X,{\rm End}(E))$.  Likewise, by arguments similar to those
for the $\A$ model, $L_0$ and
$b_0$ reduce to
$$\eqalign{L_0 & = {\pi\over 2 t}\left(\overline\partial\,\overline\partial^*
+\overline\partial^*\overline\partial\right) \cr
           b_0& = {\pi\over 2t}\overline\partial^*.\cr}\eqn\ovoco$$

\chapter{Space-Time Interpretation Via String Field Theory}

In this section we will determine the space-time interpretation of the
models just constructed in the case of open strings, that is in the case
of orientable world-sheets with boundary.  The reason for focussing on this
case is that a simple and beautiful answer will arise, as we will see.
The more perplexing closed string case (Riemann surfaces without boundary)
is the subject of \S5.

Usually, closed strings are inescapable even when one tries to do open
string physics, for elementary topological reasons.  The reader may
therefore be surprised at the lengthy discussion of open strings in this
section ignoring closed strings.  In fact, as we discuss in \S4.2,
our main points can be formulated at the classical level, where closed
strings cannot appear; also, we argue in \S5 that in these topological
theories, the open and closed strings are decoupled.

%%%%%%%%%%%%%%%%%%%%%%%%%%%%%%%%%
%Insert Figure 3  csg3Xa-b.eps
\epsfxsize=3.5in
\topinsert\centerline{\epsfbox{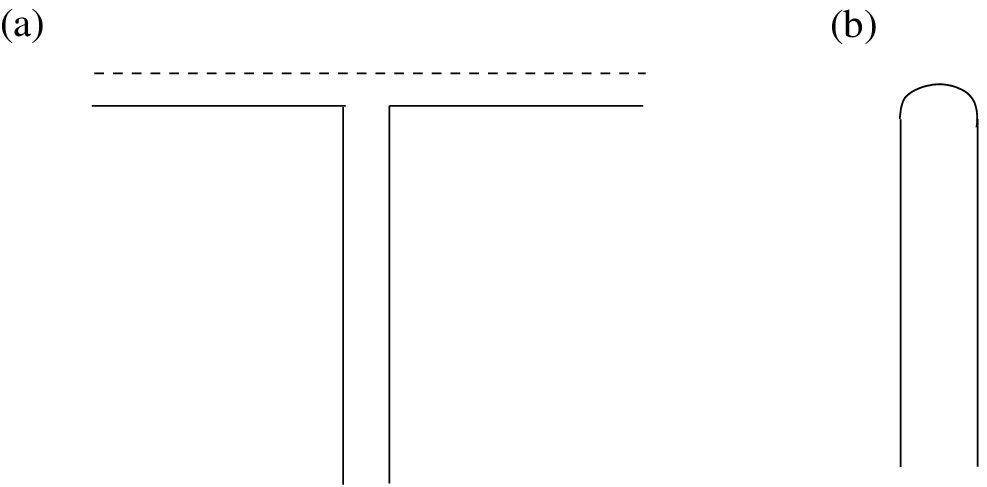}}
{\narrower\smallskip\singlespace
\noindent Fig. 3.
The multiplication law (a) and integration law (b) of
open string field theory are defined by pictures involving gluing of
strings.
\smallskip} \endinsert
\noindent

\FIG\multo{}

%\FIG\multo{The multiplication law (a) and integration law (b) of
%open string field theory are defined by pictures involving gluing of
%strings.}
The simplicity of the
open string topological theories
is closely related to the comparative simplicity
of open string field theory [\ewitten], which in turn
is intimately tied up [\martinec] with the existence of a simple
cell decomposition of the moduli space of Riemann surfaces with boundary
[\penner--\other].  This cell decomposition
involves the construction of Riemann surfaces by gluing together flat
strips $S_i$ of width $\pi$ and length $T_i$; they are glued, as sketched
in figure (\ulto),  at vertices
that generically (unless some $T_i=0$) are trivalent.  The resulting
pictures have been called ``fatgraphs'' by Penner.  In the limit
when the $T_i$ are all large, or equivalently when the widths of the strips
can be neglected, the fatgraphs of the open-string field theory reduce
to ordinary graphs corresponding to Feynman diagrams of ordinary field theory.

\subsection{Construction Of Open String Field Theory}

We recall now the highlights of the construction of open string field theory.
A key part of the structure is a
$\IZ$-graded associative algebra ${\cal B}$,
with a multiplication law
that we denote as $\star$, and a derivation $Q$ of degree $1$ with
$Q^2=0$.
The degree is usually called ghost number.\foot{The ghost
number as I will count it here
is the ghost number as defined in [\ewitten] plus $3/2$; thus,
the $SL(2,\IR)$ invariant vacuum has ghost number
zero.}
There is a linear functional $\int:{\cal B}\to\IC$, of degree
$-3$ (that is, if $b$ is an element of ${\cal B}$ of definite ghost number,
then $\int b$ vanishes unless $b$ has ghost number three), obeying
$\int a\star b=(-1)^{\deg a\deg b}\int b\star a$, $\int Qb=0$,
for all $a,b\in {\cal B}$.  The string field is a ghost number 1 element
${\cal A}\in {\cal B}$.  The Lagrangian is
$$L={1\over 2}
\int\left({\cal A}\star Q{\cal A}+{2\over 3}{\cal A}\star {\cal A}
\star {\cal A}\right). \eqn\umbo$$
This is invariant under gauge transformations generated by
$$\delta {\cal A}= Q\epsilon -\epsilon\star {\cal A}+{\cal A}\star \epsilon.
        \eqn\ubbo$$

Chan-Paton factors are introduced as follows (in the case of a trivial
rank $N$ bundle in space-time with flat connection; the generalization
is discussed presently).  Let $M_N(\IC)$ be
the associative algebra of $N\times N$ complex matrices.  One simply
replaces ${\cal B}$ with ${\cal B}\otimes M_N(\IC)$ (and $\int$ by
$\int\otimes \Tr$, $\Tr$ being the usual trace on $M_N(\IC)$).
This preserves the basic structures.  In this process, ${\cal A}$ acquires
matrix indices.  If a suitable reality condition is imposed, ${\cal A}$
takes values in $N\times N$ hermitian matrices -- the Lie algebra of
$U(N)$.

Conventionally, ${\cal B}$ is taken to be the space of open string
states in some critical string theory; the multiplication and integration
operations $\star$ and $\int$ are defined by operations of gluing strings
that we recall in figure (\multo). $Q$ is the BRST operator of the critical
string theory.  The integration law is of ghost number $-3$ since the Euler
characteristic of a disc is $-1$, and in critical string theory the
ghost number of the vacuum is $-3\chi(\Sigma)$.

Now we want to consider open string field theory with the conventional
string models replaced by the more exotic ones discussed in \S2-3.
In particular, for the world-sheet theory we take a topological sigma
model with a Calabi-Yau target space $X$ of complex dimension three; this
ensures the correct ghost number of the vacuum -- or of the integration law.
Multiplication and integration are defined by the standard gluing
operations; $Q$ will now be the BRST operator of whatever topological
field theory we consider.  This framework enables us to construct
an open-string field theory from any world-sheet theory (obeying the
usual axioms).  For instance, instead of tensoring with $M_N(\IC)$ to introduce
trivial Chan-Paton factors,
we can use the world-sheet theories with boundary interactions
constructed in \S3.2 to couple to a flat bundle on $M$ in the case of
the $\A$ model, or a holomorphic bundle on $X$ in the case of the $\B$ model.

\section{The $\A$ Model}

We want to understand the physical content of these models.
First we consider the $\A$ model.
It is necessary to pick boundary conditions, and as in \S3 we follow
Floer and pick boundary conditions associated with Lagrangian submanifolds
of $X$.  In \S3, in studying a particular surface $\Sigma$, we
introduced a separate Lagrangian submanifold $M_i$
for each boundary component $C_i$ of $\Sigma$.  In string field theory,
one generates all possible $\Sigma$'s via a Feynman diagram expansion,
and the $M_i$'s must be built in universally at the outset.  We will do
this in the simplest way by picking a single $M$ once and for all.
Thus the free boundaries of all our strings and surfaces will be mapped to
$M\subset X$.
(Generalizations exist; by correlating the choice of
$M$ with the Chan-Paton factors one could make a gauge invariant string
field theory with more than one $M$.)

Since $X$ is of complex dimension three, $M$ is a three-manifold.
A neighborhood of $M$ in $X$ is equivalent topologically and even
symplectically to a neighborhood of $M$ in its cotangent bundle $T^*M$.
The topological string theory with target space $X$ involves, roughly,
two ingredients.  One is the instantons with target $X$ (and boundary values
in $M$); these are the usual subjects of study in Floer theory.
The other side of the story, as we will see, involves Chern-Simons theory
with target space $M$.  I want first to isolate this
``new'' ingredient -- new in the sense
that it is not usually coupled with Floer theory.
The instantons can be suppressed and the new ingredient isolated by
replacing $X$ by $T^*M$, since a simple vanishing theorem (discussed presently)
shows that there are no non-constant instantons mapping $\Sigma$ to $T^*M$.
Later we will generalize to arbitrary $X$ and determine the instanton
corrections to the space-time Chern-Simons theory.

So until further notice, in discussing the $\A$ model, our target space will be
the cotangent bundle $T^*M$ of an oriented three-manifold $M$.
\foot{The orientation is needed to consistently define the sign of the fermion
determinant in the world-sheet theory of \S3.}
Like any symplectic manifold, $T^*M$ can be given an almost complex
structure such that the symplectic structure is positive and of type $(1,1)$;
indeed, this is essential in Floer/Gromov theory of symplectic manifolds.
This is good enough for formulating the ${\bf A}$ model with target
space $T^*M$; though the
transformation laws and Lagrangian of the $\A$ model were written
in \S3 in a way that assumed the integrability of the almost complex structure,
this assumption can be relaxed, as explained in detail in [\topsig].

\subsection{The Vanishing Theorem}

Now let us briefly explain the vanishing theorem, which asserts
that instantons mapping $\Sigma$ to $T^*M$ and mapping $\partial\Sigma$
to $M$ are necessarily constant.

Consider temporarily a general symplectic manifold $X$ with symplectic form
$\omega$ and an almost complex structure $J$ such that $\omega$ is
of type $(1,1)$ and positive.  Positivity means that the metric $g_{IK}
=J^S{}_I\omega_{SK}$ is positive definite; the $(1,1)$ condition
means that if $i,j,k$ and $\overline i,\overline j,\overline k$ are
indices of types $(1,0)$ and $(0,1)$, respectively, then
$g_{ij}=g_{\overline i\overline j}=0$, and $g_{i\overline j}
=g_{\overline j i}=-i\omega_{i\overline j}$.
If $\Sigma$ is a Riemann surface, an instanton or almost holomorphic
map is a map $\Phi:\Sigma\to X$ with $\bar\partial \phi^i=0$.  Consider
the bosonic sigma model action
$$I= i\int_\Sigma \d z\wedge \d\overline z \,\,\,\,\,g_{IJ}\partial_z\phi^I
\partial_{\overline z}\phi^J.   \eqn\kufo$$
Instantons minimize this action for a given homotopy class
since
$$\eqalign{ I= &
2i\int_\Sigma\d z\wedge \d\overline z \,\,\,\,\,g_{\overline i j}\partial_z
\phi^{\overline i}\partial_{\overline z}\phi^j
-i\int_\Sigma \d z\wedge \d\overline z \,\,\,\,\,g_{\overline i j}\left(
\partial_z\phi^{\overline i}\partial_{\overline z}\phi^j
- \partial_z\phi^{j }\partial_{\overline z}\phi^{\overline i}\right)\cr
=& 2i\int_\Sigma\d z\wedge \d\overline z \,\,\,\,\,g_{\overline i j}\partial_z
\phi^{\overline i}\partial_{\overline z}\phi^j
+\int_\Sigma\Phi^*(\omega). \cr}
       \eqn\jufo$$
The first term on the right hand
on the right hand side of \jufo\ is positive semi-definite and vanishes
precisely for instantons; so for instantons, the action reduces to
$I=\int_\Sigma\Phi^*(\omega)$.  The vanishing theorem comes by showing
that if $X=T^*M$, and $\Phi(\partial\Sigma)\subset M$, then
$$ \int_\Sigma \Phi^*(\omega) = 0.     \eqn\ncnc$$
If this is known, then $I$ vanishes for instantons that map $\partial\Sigma$
to $M$; but from the definition \kufo\ it is clear that $I$ vanishes
only for constant maps.

To justify \jufo, pick on $M$ local coordinates
$q^a, \,\,\,a=1\dots 3$.
The symplectic structure of $T^*M$ can be written
as $\omega =\sum_{a=1}^3\d p_a\wedge\d q^a$, with $p_a$
linear coordinates in the fibers that vanish on $M$.
This is $\omega=\d \rho$, where $\rho=\sum p_a \d q^a$
vanishes on $M$.
So $\int_\Sigma\Phi^*(\omega)=\int_{\partial\Sigma}\Phi^*(\rho)=0$,
if $\Phi(\partial\Sigma)\subset M$.

\subsection{Low Energy Expansion}

As explained following equation \ddalfo,
the key simplification of the $\A$ model is that the essential physics
is independent of the coupling parameter $t$.  As $t$ and the target
space metric $g_{IJ}$ appear only in the combination $tg_{IJ}$,
large $t$ is simply the limit in which the target space metric is scaled
up; it is the limit of large distances or long wavelengths.
This is the limit
in which ordinary string theory reduces approximately to field theory.
Since ordinary string theory is $t$-dependent, the large $t$ behavior is
only an approximation.  The topological string theories that we are studying
are $t$-independent, so one can hope for an exact description by looking
at the large $t$ behavior.

We have analyzed the large $t$ behavior of the string states in \S3.3.
In particular, as we saw there, the low-lying modes can be described
as functions ${\cal A}(q^a,\chi^b)$ of the center of mass variables.
In \S3.3, we computed the cohomology for all ghost numbers, but
in the string field theory context, we want to insist that
that ${\cal A}$ have ghost number 1.  This  means that
it must be linear in $\chi$, so we write
$${\cal A}=\chi^a A_a(q), \eqn\nunu$$
with $A_a\d q^a$ a one-form on $M$, valued possibly in ${\rm End}(E)$,
the endomorphisms of a flat vector bundle $E$.

In \S3.3, we identified $Q$, in the same large $t$ limit,
with the exterior derivative $\d$.  Then the quadratic part of the
string field Lagrangian ${1\over 2}\int {\cal A}\star Q{\cal A}$
reduces (for the
low-lying modes in the large $t$ limit) to
$${1\over 2}\int_M\Tr A\wedge \d A, \eqn\umigo$$
with $\Tr$ a trace on the Chan-Paton factors.
This is the quadratic part of the ordinary Chern-Simons action for
a gauge field $A$.

%%%%%%%%%%%%%%%%%%%%%%%%%%%%%%%%%
%Insert Figure 4  csg4X.eps
\epsfxsize=1.5in
\topinsert\centerline{\epsfbox{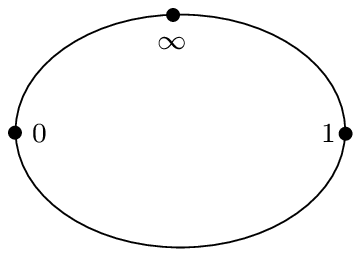}}
{\narrower\smallskip\singlespace
\noindent Fig. 4.
The coupling of three gauge fields on a disc.
\smallskip} \endinsert
\noindent
\FIG\disco{}

%\FIG\disco{The coupling of three gauge fields on a disc.}
What about the cubic part of the string field action?
If $A^{(j)}$ is a mode of the gauge field $A$, the corresponding
vertex operator is $V^{(j)}=
\chi^a A^{(j)}_a(q)$.  Let us evaluate
by world-sheet path integrals the coupling
of three such modes on the
disc.  This is done by inserting three vertex operators
$V^{(i)},\,\,i=1\dots 3$
at boundary points of the disc, as in figure (\disco).
Because of $SL(2,\IR)$ symmetry, all that
matters in the choice of the three points is the cyclic order.
So we can take $\langle V^{(1)}(0)V^{(2)}(1)V^{(3)}(\infty) \rangle$.
In the large $t$ limit, the path integral reduces to an integral over
the constant zero modes; this integral gives
$$\int \d q^1\dots \d q^3 \,\d\chi^1\dots \d\chi^3
  \Tr\chi^a A^{(1)}_a(q)\,\,\chi^b A^{(2)}_b(q)
  \,\, \chi^c A^{(3)}_c(q)=\int_M\Tr A^{(1)}\wedge A^{(2)}\wedge A^{(3)}
\eqn\untangle$$
This is simply a matrix element of the cubic part of the usual Chern-Simons
action.  So putting the pieces together,
the string field action \umbo\ reduces to the ordinary Chern-Simons
Lagrangian
$$L_T={1\over 2}\int_M\Tr\left(A\wedge\d A+{2\over 3}A\wedge A\wedge A\right).
\eqn\dotangle$$
The string field gauge invariances can similarly be seen in the same
approximation to reduce to conventional gauge invariance.
So, under the conditions that we have formulated, the abstract Chern-Simons
Lagrangian of string field theory reduces to the ordinary Chern-Simons
Lagrangian.

Now usually, string theories reduce to field theories at low energy
only approximately.  For instance, ordinary string
theories reduce approximately at low energies to general relativity plus
massless matter fields, but there are $O(\alpha')$ corrections.

In the case of topological string theories (such as the $\A$ model
considered here) the situation is quite different.  The analog of
the usual $O(\alpha')$ corrections
would be perturbative corrections in $1/t$.
These are absent, because the model is $t$-independent.  Instanton
corrections are also absent, because for target space $T^*M$, there are
no non-constant instantons, as we have seen.
So the reduction of the $\A$ model to \dotangle\ is exact.
If we replace $T^*M$ by some Calabi-Yau threefold $X$ (having $M$ as a
Lagrangian subspace), there are instanton corrections
to \dotangle\ that we will discuss later.

\section{Reconsideration From The Hamiltonian Point Of View}

Now I want to reconcile what we have said above with previous discussions
of the $\A$ model such as [\mirror] (or [\topsig,\S4.1] where the coupling
to topological gravity is considered).

Formally, the $\A$ model is a quantum field theory that
counts  holomorphic curves obeying various conditions.
(This is familiar in applications of the $\A$ model
to mirror symmetry, for instance.)
However, in the vanishing theorem that we have discussed above,
we have seen that for target space $T^*M$, there are no non-constant
instantons.
How therefore can the model possibly be non-trivial, and in particular
equivalent as we have claimed to perturbative Chern-Simons gauge theory?

\REF\mathai{V. Mathai and D. Quillen, ``Superconnections, Thom Classes,
and Equivariant Differential Forms,'' Topology {\bf 25} (1986) 85.}
\REF\atj{M. F. Atiyah and L. Jeffrey, ``Topological Lagrangians
And Cohomology,'' J. Geom. Phys. {\bf 7} (1990) 119.}
To answer this question, it is necessary to recall some background.
Formally, in topological quantum field theories of this general nature
(``cohomological field theories''),
one has an infinite dimensional function space ${\cal S}$, and over
it an infinite dimensional vector bundle ${\cal W}$; one wishes
to make sense of the Euler class of this bundle.
To tame the situation, one has a section $w$ of ${\cal W}$.  In finite
dimensions, if ${\cal S}$ is compact and without boundary,
the Euler class of ${\cal W}$
would be Poincar\'e dual to a homology class $[w]$ supported on the zeros
of $w$.  The Euler class of ${\cal W}$, integrated over ${\cal S}$,
can be represented by various integral formulas; one can construct
an integral formula which is localized near the zeros of $w$ [\mathai].
The Feynman
path integral of a cohomological field theory is a function space
version of such a formula.  This point of view is developed
(in an analogous four dimensional gauge theory) in [\atj].

To identify ${\cal S}$, ${\cal W}$, and $w$ in the case at hand,
let $\Sigma$ be a compact oriented two dimensional surface, perhaps
with boundary, not endowed with any {\it a priori} complex structure.
Let $X$ be a symplectic manifold, with an almost complex structure
obeying the usual conditions, and with a Lagrangian submanifold $M$.
Fix a homotopy class ${\cal H}$
of maps of $\Phi:\Sigma\to X$  with $\Phi(\partial\Sigma)\subset M$.
Such a map can be described locally by complex functions $\phi^i$ on
$\Sigma$ (corresponding to local complex coordinates on $X$).
Then we set ${\cal S}$
to be the space of pairs $(J,\Phi)$, where $J$ is a complex
structure on $\Sigma$, and $\Phi$ is a map $\Sigma\to X$ in the class
${\cal H}$, with two such pairs considered equivalent if they differ
by a diffeomorphism of $\Sigma$.  For ${\cal W}$ we take the vector
bundle over ${\cal S}$ whose fiber is the space of
sections of $\Omega^{0,1}(\Sigma)\otimes \Phi^*(T^{1,0}X)$, where
$\Omega^{0,1}(\Sigma)$ is the space of $(0,1)$-forms on
$\Sigma$ in the complex structure $J$, and $T^{1,0}X$ is the holomorphic
tangent bundle of $X$.
For $w$, we pick a natural section of ${\cal W}$ given by the equation
for an almost holomorphic curve; it can be written
$w=\overline\partial
_J\phi^i$, with $\overline\partial_J$ the $\overline\partial$ operator
on $\Sigma$ determined by $J$.

Now, let us focus on the case $X=T^*M$, endowed with an almost complex
structure of the usual type and so in particular with a metric.
The vanishing theorem discussed earlier implies that for $H$ a non-trivial
homotopy class, there are no instantons; $w$ has no zeros.
If ${\cal S}$ were compact, we would conclude that the Euler class of
${\cal W}$ is zero.

%%%%%%%%%%%%%%%%%%%%%%%%%%%%%%%%%
%Insert Figure 5  csg5-X.eps
\epsfxsize=4in
\topinsert\centerline{\epsfbox{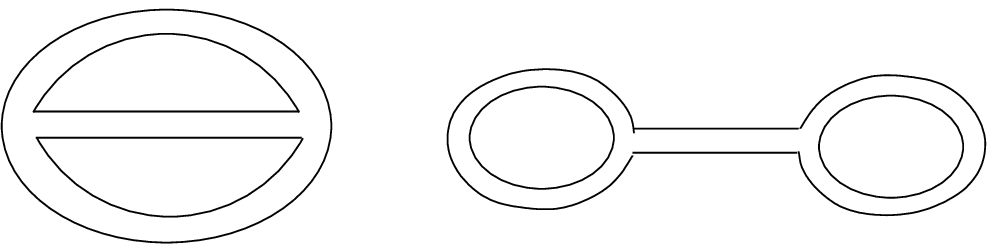}}
{\narrower\smallskip\singlespace
\noindent Fig. 5.
Degeneration of a three-holed sphere to a trivalent graph can occur in two ways.
\smallskip} \endinsert
\noindent
\FIG\degenerate{}

%\FIG\degenerate{Degeneration of a three-holed sphere to a trivalent
%graph can occur in two ways.}
However, ${\cal S}$ is non-compact, and its non-compactness is essential.
In fact,
in a certain sense, there are instantons or zeros of $w$ at infinity.
These arise
in a limit in which the Riemann surface degenerates to a graph $\Gamma$.
This possibility is sketched in figure (\degenerate); from the point
of view of open-string field theory, the degeneration arises
in the limit in which all $T_i$ go to infinity in figure
(\ulto).

In a given homotopy class of maps of the graph
$\Gamma$ to $X$, there will be a map
$\Phi$ for which $\Phi(\Gamma)$ is a ``geodesic graph'' of least total length.
$\Phi$ can be regarded as a limit of maps from almost degenerate Riemann
surfaces (obtained by slight thickening of $\Gamma$) that almost obey the
instanton equation.  It is an ``instanton at infinity.''
The sigma model action \kufo\ (which should vanish for instantons obeying
our boundary conditions, as we saw in the proof of the vanishing theorem)
vanishes in the limit of such a geodesic graph.

\REF\fried{D. Fried, Inv. Math. {\bf 84} (1986) 523.}
The existence of zeros at infinity means that there is no straightforward
evaluation of the world-sheet path integral by counting zeros of $w$.
There are no real zeros, but there are virtual zeros at infinity.
While the contributions of real zeros are integers that can be counted
topologically, there is no apparent reason for this to be true for the
virtual zeros at infinity.
In any event, the contributions of the virtual zeros, summed over all homotopy
classes, can be evaluated by computing
the measure on the moduli space (of complex structures on $\Sigma$) that
comes from the world-sheet path integral.
\foot{When $\pi_1(M)$ has infinitely many conjugacy
classes, so that the number of homotopy classes is infinite, it is scarcely
plausible that the full answer can be written as a sum over homotopy classes
with each class contributing an integer.  For convergence,
the contributions would have to
be almost all zero.  The possible
noncommutativity of
the sum over homotopy classes and the integration over moduli space may
also mean that the final answer (which should be obtained by summing over
homotopy classes first) cannot be written as a sum of contributions of
homotopy classes, though the measure on moduli space can be so written.
In this case, the question of whether the individual contributions are
integers does not even make sense.
This seems to be the case for the one loop contributions, as one can see by
comparing to D. Fried's formulas expressing the logarithm of
analytic torsion
as a sum over geodesics [\fried].}

The essential simplification of the $\A$ model always arises
because  for large $t$, this measure is concentrated near zeros of $w$.  In
the case that the only zeros are virtual zeros at infinity, the measure is
concentrated for large $t$ on the region near infinity; but this
is the limit in which string theory
reduces to field theory, and world-sheet path integrals reduce to Feynman
graphs.
So the non-vanishing contributions to the world-sheet path integral
come entirely from the regions in the moduli space (of complex structures
on $\Sigma$) in which the string
theory reduces to the evaluation of Feynman graphs.
This is the equivalence of the string theory to a field theory
that was claimed earlier.
Of course, one must remember that
 (as in the figure, where there are two possible
degenerations) one Riemann surface will in general be capable of
degenerating to several possible Feynman graphs.

\subsection{Perturbation Theory}

To make this a little more explicit, we will now discuss how to extract
the field theoretic Feynman rules from the string theory.

\REF\thorn{C. B. Thorn, ``Perturbation Theory For Quantized String
Fields,'' Nucl. Phys. {\bf B287} (1987) 61.}
\REF\bocc{M. Bochicchio, ``Gauge Fixing For The Field Theory Of The
Bosonic String,'' Phys. Lett. {\bf 193B} (1987) 31.}
We want to describe the measure on the moduli space of Riemann surfaces
that follows from the description of moduli space via fatgraphs.
This can be worked out systematically by quantizing open string
field theory.  The necessary techniques were given by Thorn and Boccicchio
[\thorn,\bocc] and involve an elegant application of the Batalin-Vilkovisky
approach to quantization.  Instead of following that road, I will merely
use the fatgraph description of moduli space and the definition of the
measure in equation \cocn.

%%%%%%%%%%%%%%%%%%%%%%%%%%%%%%%%%
%Insert Figure 6  csg6X.eps
\epsfxsize=2.5in
\topinsert\centerline{\epsfbox{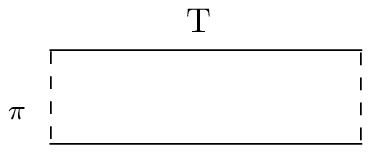}}
{\narrower\smallskip\singlespace
\noindent Fig. 6.
A strip $S$ of width $\pi$ with one real modulus $T$.
\smallskip} \endinsert
\noindent

\FIG\strip{}
%\FIG\strip{A strip $S$ of width $\pi$ with one real modulus $T$.}
The basic fact that we will use is therefore the fact that every Riemann
surface $\Sigma$ with boundary has a canonical flat metric (with some conical
singularities) built as in figure (\ulto).
This metric is obtained by gluing together some standard strips.
The strips are flat rectangles of width $\pi$ in the $\sigma-\tau$ plane, say
$0\leq\sigma\leq \pi$, $0\leq\tau\leq T$,
with metric $\d s^2=\d\sigma^2+\d\tau^2$.  Such a strip, say $S$ (figure
(\strip)),
has one real modulus,
namely $T$.  A deformation of this modulus from $T$ to $T+\delta T$
can be made by changing the metric to
$$\d s^2=\d\sigma^2+\d\tau^2\left(1+2f(\tau)\right), \eqn\noro$$
with $f(\tau)$ any function with $\int_0^T\d\tau f(\tau)=\delta T$.
It is convenient to take $f(\tau)=\delta T\cdot \delta (\tau-\tau_0)$,
for some $\tau_0$.  If one of the $\delta^{(k)}h$ in equation \cocn\ is the
deformation just described of the length of the strip $S$, then the
corresponding mode $b^{(k)}$ of equation \gogo\ is
simply
$$b^{(k)}=\int_0^\pi\d\sigma \,\,\,b_{00}(\sigma,\tau_0). \eqn\ncnc$$
This is in other words an insertion of the zero mode $b_0$ at $\tau=\tau_0$.

We also have to worry about the propagation of the string through the
proper time $T$; this gives a factor of $\exp(-TL_0)$, with $L_0$ the
string Hamiltonian discussed in \S3.3.  Allowing for the integral over $T$
and the insertion of $b_0$, we get for each strip a factor
$$\int_0^\infty \d T  \,\,b_0 \exp(-TL_0). \eqn\kici$$
Let ${\cal V}$ be the kernel of $L_0$, and let $1/L_0$ be the inverse
of $L_0$ in the subspace orthogonal to ${\cal V}$.  Since $b_0$
annihilates the kernel of $L_0$,\foot{The topological sigma model
in question was constructed by twisting an ordinary supersymmetric
model (as explained in detail in [\mirror]) by a power of the canonical
bundle.  On the flat strip under discussion here, the canonical bundle
is naturally trivialized, and the twisting does nothing.  In the underlying
supersymmetric model, $b_0$ is the adjoint of $Q$, and as in Hodge
theory, the relation
$L_0=\{Q,b_0\}$ implies that $Q$ and $b_0$ annihilate the kernel of $L_0$.}
the integral in \kici\ can be evaluated to give
$$\int_0^\infty \d T  \,\,b_0 \exp(-TL_0)=b_0{1\over L_0}.\eqn\omob$$

In equations \ungo\ and \nugo,
we determined the large $t$ limits of $L_0$ and $b_0$ to be $\pi\Delta/2t$
and $\pi\d^*/2t$.  (Thus, $b_0$ vanishes for large $t$, but $b_0/L_0$ has
a non-vanishing limit because of the small eigenvalues of $L_0$.)  So
the large $t$ limit of \omob\ is
$$ \d^*{1\over \Delta}.     \eqn\ymob$$
Now, \omob\ is the open string propagator, while \ymob\
is the propagator of Chern-Simons field theory.

In the fatgraph approach to construction of moduli space, the strips
we have just analyzed are glued together at cubic vertices to make
Riemann surfaces.  The structure of the interaction vertices is the same as in
the open string multiplication law, figure (\multo(a)).
We have already seen in connection with figure (\disco) how these
vertices reduce at large $t$ to wedge products of differential forms,
as in Chern-Simons field theory.

While the underlying string field ${\cal A}$ had ghost number 1,
and the underlying Chern-Simons gauge field $A$ is a one-form,
in \omob\ and \ymob\ states of all ghost number -- or differential forms
of all degrees -- are propagating.  In either case, the vertices of the gauge
fixed theory have in a natural sense the same structure (gluing of strings
or wedge products) as that in the classical theory, though the constraint
on the ghost number or degree is absent.
In the case of string theory, this structure was explained by Thorn
and Boccicchio [\thorn,\bocc] by solving the master equation and
gauge fixing of string field theory.
In Chern-Simons theory, the analogous formulation of
the gauge fixed theory, combining the fields of different ghost number,
is due to Axelrod and Singer [\axelrod].

\subsection{The Tree Level $S$-Matrix}

%\FIG\newdisco{Coupling of four states $\alpha_{(1)},\dots,\alpha_{(4)}$ on
%the boundary of a disc (a) with an antighost insertion
%around $\alpha_{(2)}$; the Feynman diagrams corresponding
%to the two possible ``channels'' (b).}
I would now like to make this more explicit in a particular example.
I have picked the example to anticipate the following question that
may perplex some readers.  We have extracted Chern-Simons theory from
the open string degenerations of moduli space, but are there additional
closed string contributions?  It will be clear in \S5 that the role of the
closed strings is not fully understood, though I will argue there that
the closed and open strings are decoupled.
I therefore want to demonstrate that
the statement ``the topological string theory of the $\A$ model is
equivalent to Chern-Simons field theory'' can be tested in a situation
in which closed strings are definitely not relevant.  This is the case
for scattering amplitudes (or the analog of what in ordinary string theory
would be scattering amplitudes) for the case that $\Sigma$ is a disc.

%%%%%%%%%%%%%%%%%%%%%%%%%%%%%%%%%
%Insert Figure 7  csg7Xa-b.eps
\epsfxsize=4in
\topinsert\centerline{\epsfbox{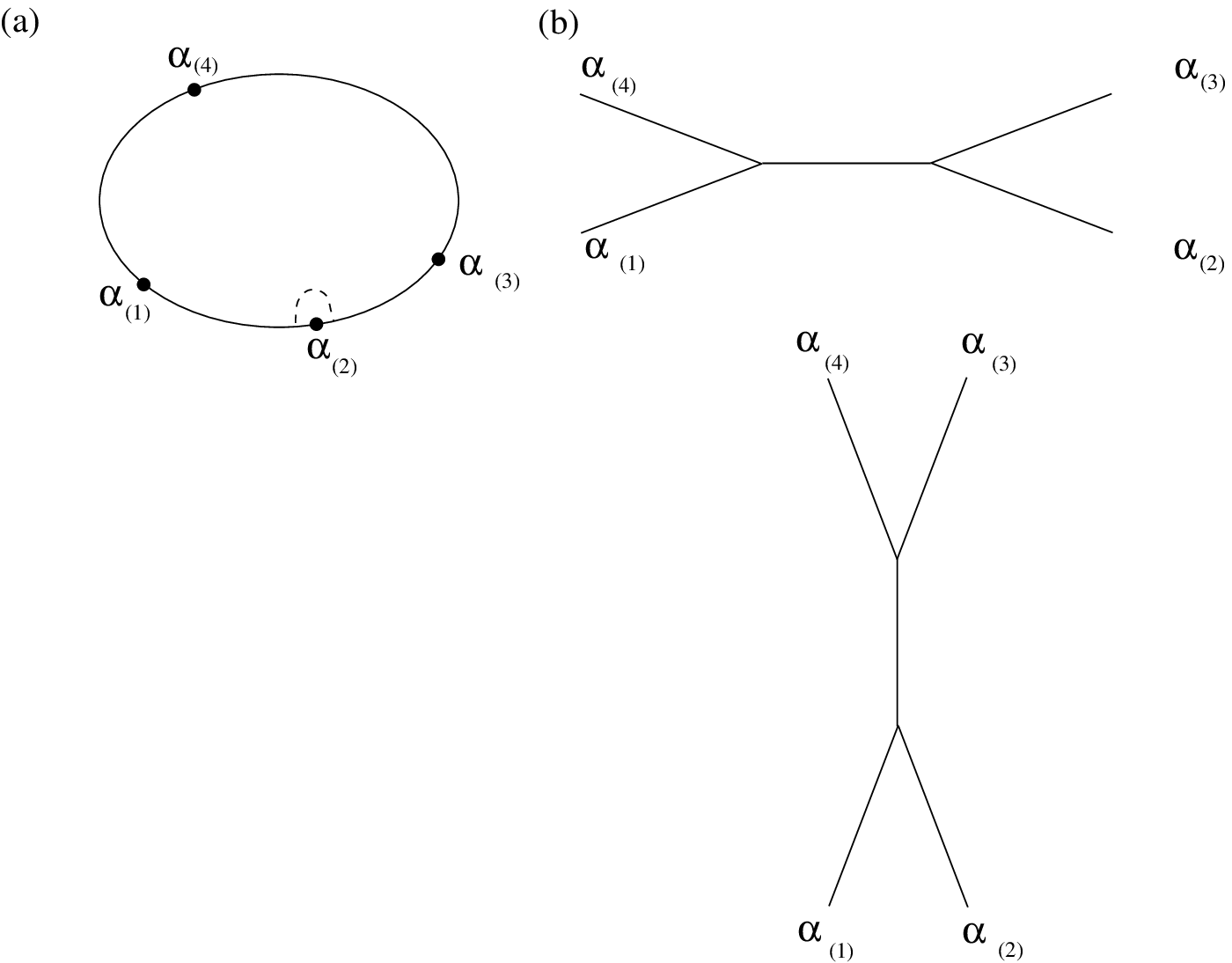}}
{\narrower\smallskip\singlespace
\noindent Fig. 7.
Coupling of four states $\alpha_{(1)},\dots,\alpha_{(4)}$ on
the boundary of a disc (a) with an antighost insertion
around $\alpha_{(2)}$; the Feynman diagrams corresponding
to the two possible ``channels'' (b).
\smallskip} \endinsert
\noindent

\FIG\newdisco{}
As we know from \S3.3, the ``physical states'' of the open string
field theory are elements of $H^1(\Sigma,{\rm End}(E))$, $E$ being
a flat vector bundle that we will keep fixed in the discussion.
The exterior derivative with values in that bundle will be denoted as $\d$.
Let $\alpha_{(1)},\dots ,\alpha_{(4)}$ be four such states.  We consider their
coupling on the disc in that cyclic order, as in figure (\newdisco).
To calculate the four point function in field theory, all that we really
need to know is that there are two possible Feynman diagrams, indicated
in figure (\newdisco), and that the propagator is $\d^*/\Delta$.
The amplitude is therefore
$$I(\alpha_{(1)},\dots,\alpha_{(4)})
=\int_M\Tr\left(\alpha_{(1)}\wedge \alpha_{(2)}
{\d^*\over \Delta}\alpha_{(3)}\wedge\alpha_{(4)}+\alpha_{(2)}
\wedge\alpha_{(3)}{\d^*\over\Delta}
\alpha_{(4)}\wedge\alpha_{(1)}\right).       \eqn\umuo$$

To do the same calculation in string theory, we introduce vertex operators
$V_{(i)}=\alpha_{(i)\,a}\chi^a$ for the external states, and one antighost
insertion corresponding to the one real modulus of the disc.  We can map
$V_{(1)}$, $V_{(3)}$, and $V_{(4)}$ to $0,1$, and $\infty$, and take
the modulus to be the position of $V_{(2)}$.  In that case, the antighost
insertion can be taken to be a small contour integral around the position
of $V_{(2)}$, as in the figure.
The contour integral can be evaluated by thinking of $V_{(2)}$
as an operator inserted on the end of an open string in a Hamiltonian
formulation; the antighost contour computes the commutator of
this operator with the antighost zero mode $b_0$,
and so replaces the zero form
operator $V_{(2)}$ by its one-form descendant
$$W_{(2)} = [b_0, V_{(2)}] = A_a(q){\d q^a\over \d \tau}.    \eqn\muo$$
The string theory amplitude is therefore
$$\int_0^1\d\sigma \langle V_{(1)}(0) \cdot A_a{\d q^a\over \d\tau}
     (\sigma)\cdot
   V_{(3)}(1) V_{(4)}(\infty) \rangle.  \eqn\uo$$
Now we wish to evaluate this, as usual, in the limit of $t\to\infty$.
If the integral were limited to a range of $\sigma$ bounded away from
$0$ and 1, it would vanish as $1/t$ for $t\to \infty$.  This is so because
the $\phi$ and $\psi-\chi$ propagators are both proportional to $1/t$.
A non-zero result arises only because if the result is blindly expanded
in powers of $1/t$, the endpoint contributions are infinite.  This is
because the term of lowest order in $1/t$
in the operator product $V_{(1)}(0)\cdot A_a{\d q^a\over \d\tau}(\sigma)$
is proportional to $1/\sigma$, coming from $\phi-\partial\phi$ contractions.
A simple-minded $1/t$ expansion of \uo, evaluating those contractions
(and discarding a term proportional to $\{Q,\dots\}$),
leads to a divergent contribution near $\sigma=0$ that looks like
$${\pi \over 2t}\int_0^{\epsilon}\d\sigma {1\over \sigma}
\langle \chi^a\left(\d^*(\alpha_{(1)}\wedge\alpha_{(2)})\right)_a(0)\,\,\,
 V_{(3)}
(1)\,\,\,\,\,V_{(4)}(\infty)\rangle,
             \eqn\noggo$$
with $\epsilon$ an arbitrary fixed positive number.

To cure the infinity,
we must remember the anomalous dimensions of the various operators, which
are given by the eigenvalues of $L_0$.
To order $1/t$, we had $L_0=\pi\Delta/2t$ in \S3.3.
The physical states have harmonic representatives, and if we pick such
representatives (discarding terms of the form $\{Q,\dots\}$), the
$V_{(i)}$ have $L_0=0$ in order $1/t$.   However, the vertex operator
$\chi^a\left(
d^*(\alpha_{(1)}\wedge\alpha_{(2)})\right)_a$ in \noggo\ corresponds to
a state not annihilated by $L_0$.  Including powers of $\sigma$ coming
from the anomalous dimensions, and
replacing the expectation values $\langle ~~~~\rangle $ by zero mode
integrals,
\noggo\ is replaced by
$${\pi\over 2t}\int_0^{\epsilon}\d\sigma {1\over \sigma}
\int_M\Tr\left(\d^*(\alpha_{(1)}\wedge\alpha_{(2)}) \sigma^{\pi\Delta/2t}
\wedge \alpha_{(3)}
\wedge \alpha_{(4)}(\infty)   \right)\eqn\noggo$$
The endpoint contribution of the $\sigma$ integral is now finite as $t\to
\infty$, and reproduces one of the two Feynman diagram contributions in
equation \umuo.  The other arises from the endpoint at $\sigma=1$.

Now let us discuss briefly the physical significance of these couplings.
Given a flat bundle $E$ corresponding to a solution $A_{(0)}$ about
which one wishes to expand, the space of linearized deformations is
$H^1(M,{\rm End}(E))$.  Writing $A=A_{(0)}+B$, impose the gauge condition
$\d^* B= 0 $ (with $\d^*$ being defined with respect to $A_{(0)}$).
In turn, write $B=\sum_i u_i \alpha_{(i)}+\sum_\sigma v_\sigma \beta_{(\sigma)}
$, where $\alpha_{(i)}$ are a basis of zero modes of the Laplacian $\Delta$
and $\beta_{(\sigma)}$
are the non-zero modes; and $u_i,v_\sigma$ are real coefficients.
The underlying Lagrangian
$$L=\int_M\Tr\left(A\wedge \d A+{2\over 3}A\wedge A\wedge A\right) \eqn\ico$$
can be regarded  as a function $L(u_i,v_\sigma)$.  The tree diagrams
that we have discussed above are a calculational technique for
``integrating out'' the
$v_\sigma$ to get an ``effective potential'' $V(u_i)$.
It can be written in terms of the $n$-point functions on the disc
discussed above for $n=4$ as
$$V(u_i)=\sum_{i,j,k}{u_iu_ju_k\over 3!}\langle \alpha_{(i)}\alpha_{(j)}
\alpha_{(k)}\rangle +\sum_{i,j,k,l}{u_iu_ju_ku_l\over 4!}\langle \alpha_{(i)}
\alpha_{(j)}\alpha_{(k)}\alpha_{(l)}\rangle +\dots. \eqn\ucuc$$
Since $V$ is obtained from the underlying Lagrangian $L$ by ``integrating
out the massive modes,'' its
stationary points, \ie, solutions of $\partial V/\partial u_i=0$,
are in one-to-one correspondence with the stationary points of the underlying
Lagrangian.

The process of ``integrating out the massive modes'' can be described
as follows.  Let $L$ be a function with a critical point at the origin,
and let $n$ be the dimension of the kernel of the Hessian or matrix of
second derivatives of $L$.  Then by a generalization of the Morse lemma,
it is possible to pick coordinates $u_i,i=1\dots n$
and  $v_\sigma$ in a neighborhood of the origin so that
$$L(u_i,v_\sigma)=V(u_i)+\sum_\sigma a_\sigma v_\sigma{}^2,
\eqn\orno$$
with non-zero constants $a_\sigma$; $V(u_i)$ is then the ``effective potential
with massive modes integrated out.''
The $u_i$ are of course only uniquely determined up to diffeomorphism.
Describing the possible canonical forms of $V(u_i)$ up to a reparametrization
of the variables is more or less the problem of singularity theory.

Since the $u_i$ can be described in first order as coordinates near the origin
on the
naturally defined vector space $H^1(M,{\rm End}(E))$, it is natural
to think of the effective potential $V$ as a function defined on that
space, and as such it is uniquely defined up to a coordinate transformation
$$u_i\to\widetilde u_i=u_i+c_{ijk}u_ju_k+\dots \eqn\hux$$
that is trivial in first order.
The quartic coupling $\langle \alpha_{(i)}
\alpha_{(j)}\alpha_{(k)}\alpha_{(l)}\rangle$, for particular values of
$i,j,k,$ and $l$, is invariant under such transformations
if and only if the cubic coupling $\langle \alpha_{(m)}\alpha_{(n)}
\alpha_{(p)}\rangle $ vanishes for $m,n$ equal to two consecutive values
of $i,j,k,l$;\foot{One considers $i,j,k,l$ arranged on a circle in that order,
so $l$ and $i$ are consecutive.}
by Poincar\'e duality, this is so precisely if $\alpha_{(m)}
\wedge \alpha_{(n)}$ vanishes in $H^2(M,{\rm End}(E))$
whenever $m$ and $n$ equal two consecutive values
of $i,j,k$, and $l$.

\REF\john{D. Johnson and J. Millson, ``Deformation Spaces Associated
To Compact Hyperbolic Manifolds,'' in {\it Discrete Groups In Geometry
And Analysis: Papers In Honor Of G. D. Mostow}, ed R. Howe
(Birkhauser, Boston, 1987).}
Under such conditions, the four point coupling can be given the following
interpretation.  Since $\alpha_{(i)}\wedge \alpha_{(j)}$ and
$\alpha_{(j)}\wedge \alpha_{(k)}$ vanish in cohomology,
there are ${\rm End}(E)$-valued one-forms $y,z$ with
$$\d y =\alpha_{(i)}\wedge \alpha_{(j)},~~\d z=\alpha_{(j)}\wedge\alpha_{(k)}.
             ~~\eqn\jucu$$
In fact, we can set
$$\eqalign{ y & = \d^*{1\over \Delta}\left(
                \alpha_{(i)}\wedge\alpha_{(j)}\right) \cr
            z & = \d^*{1\over
\Delta}\left(\alpha_{(j)}\wedge\alpha_{(k)}\right)
.       \cr} \eqn\numgo$$
(Because $1/\Delta$ is the inverse of $\Delta$ only in the orthocomplement
of the kernel of $\Delta$,
in proving that \numgo\ obeys \jucu\ one needs the fact
that $\alpha_{(i)}\wedge \alpha_{(j)}$ and $\alpha_{(j)}\wedge \alpha_{(k)}$
vanish in cohomology and so are orthogonal to the kernel.)
Let
$$w(\alpha_{(i)},\alpha_{(j)},\alpha_{(k)})
= y\wedge \alpha_{(k)}+\alpha_{(i)}\wedge z.   \eqn\ucux$$
Then $\d w = 0$, so $w$ determines an element of $H^2(M,{\rm End}(E))$.
Considering how $w$ transforms under a shift in $y$ and $z$, one sees
that it is not well-defined as an element of $H^2(M,{\rm End}(E))$
but only as an element of the quotient group
$$H^2(M,{\rm End}(E))/(H^1(M,
{\rm End}(E))\wedge \alpha_{(l)}\oplus \alpha_{(i)}\wedge H^1(M,{\rm End}(E))).
\eqn\ododod$$
The element $w$ of that
quotient space  is known as the Massey triple product of $\alpha_{(i)},
\alpha_{(j)}$, and $\alpha_{(k)}$.
Under the further hypothesis that $\alpha_{(k)}\wedge\alpha_{(l)}$
and $\alpha_{(l)}\wedge \alpha_{(i)}$ vanish in cohomology,
$$\int_M\Tr w(\alpha_{(i)},\alpha_{(j)},\alpha_{(k)})\wedge \alpha_{(l)}
             \eqn\hocco$$
is well-defined and in fact equals coefficient of the quartic term in the
effective potential as computed above.
The contribution of the quartic coupling to $\partial V/\partial u_l$
is therefore
$${1\over 3!}\sum_{i,j,k} u_iu_ju_k\int_M\Tr w(\alpha_{(i)},\alpha_{(j)},
\alpha_{(k)})\cdot\alpha_{(l)}.    \eqn\plocco$$
As a flat connection is a critical point of the Chern-Simons action,
this is in keeping with the observation [\john] that the Massey
triple product and its higher order analogs
are the obstructions to deformation of a flat connection.

\section{Comparison To Remarks Of Kontsevich}

What do we learn by interpreting Chern-Simons gauge theory as a string
theory?
Here I will only point out one obvious consequence.
Other applications may involve, for instance, a deeper relation
of Chern-Simons theory to mirror manifolds.

The fact that Chern-Simons theory can be interpreted as a topological
sigma model coupled to topological gravity means that one can introduce
the usual observables of topological gravity -- the stable cohomology
classes on the moduli space of Riemann surfaces introduced by Mumford,
Morita, and Miller.
In the present context, this would amount to the following.
As was sketched in \S2, the topological sigma model determines a differential
form $\Theta$ on moduli space; for open strings, the relevant moduli
space is the moduli space ${\cal M}_{g,s,h}$ of Riemann surfaces of genus
$g$ with $s$ punctures (where vertex operators are inserted) and $h$
boundary components.  Instead of integrating the closed form $\Theta$
over ${\cal M}_{g,s,h}$, we can integrate it over a homology
cycle $C$ in ${\cal M}_{g,s,h}$ of codimension, say, $n$.

The trouble is that we have fixed $X$ to have complex dimension three precisely
to ensure that $\Theta$ is a top form.
It therefore cannot be integrated over a cycle of positive codimension
unless the definition is modified somehow.
The following way of doing this will enable us to make
contact with observations by Kontsevich.

Instead of considering a single target space $X$ with a fixed
metric, we will consider a family of $X$'s with a family of metrics.
The partition function of the theory will be not a number but a closed
differential form on the space of metrics on $X$.  To implement
this idea, introduce along with the metric $g_{IJ}(\phi)$ of $X$ a fermionic
variable $\zeta_{IJ}(\phi)$ (of ghost number one) which one can think
of as (up to a constant) the exterior derivative
of $g_{IJ}(\phi)$.  Take the new transformation laws
$$\delta g_{IJ}=i\alpha\zeta_{IJ},\,\,\,\delta\zeta_{IJ}=0. \eqn\omigo$$
Recalculating the Lagrangian starting from \forlateruse, we find that
\ddalfo\ is replaced by $\widetilde L=L+L'$, with
$$L'=-t\int_\Sigma \d^2z
\zeta_{\overline i j}\left(\psi_z ^{\overline i}\partial
_{\bar z}\phi^j+\partial_z\phi^{\overline i}\psi_{\overline z}^j\right)
.\eqn\hopseo$$

Since $L'$ is of ghost number $-1$ in the ``matter fields'' $\phi,\chi,\psi$,
every insertion of $L'$ shifts the ghost number by one unit.
If we want to integrate not over moduli space but over a cycle $C$ of
codimension $n$, the non-vanishing contributions will be precisely $n^{th}$
order in $L'$.  As $L'$ is linear in $\zeta_{IJ}$,
these contributions  will be $n^{th}$ order in $\zeta_{IJ}$ and so
will define an $n$-form $\Omega$  on the space ${\cal R}$ of metrics on $X$.
$Q$-invariance means that $\Omega$ is closed.
Let ${\cal F}$ be the group of diffeomorphisms of $X$.  $\Omega$
is invariant under the natural action of ${\cal F}$ on ${\cal R}$
and moreover is basic (to show the later one notes that if
$\zeta_{IJ}=D_Iv_J+D_Jv_I$ for some $v^I$, then $L'$ is of the form
$\{Q,\dots\}$ up to terms that vanish by the equations of motion).
So if one has a fiber bundle
$$ \matrix{ X & \to & Y \cr
            \, & \, & \downarrow \cr
            \, & \, &  B, \cr} \eqn\pilx$$
with an arbitrary base $B$, then by picking a metric on the total space
$Y$ one gets a family of metrics on $X$, parametrized by $B$,
and $\Omega$ determines an $n$-dimensional cohomology class of $B$.

We have found, therefore, a map from the codimension $n$ homology of
${\cal M}_{{g,s,h}}$ to the $n$-dimensional cohomology of $B$.  This map
was described by Kontsevich [\kontsevich] by examining Chern-Simons
perturbation theory.  The considerations just explained give a more
conceptual explanation for its existence.

\subsection{A Digression}

Let us now make a small digression to examine some related observations
by Kontsevich in the light of standard quantum field theory ideas.
We want to consider the standard Fadde'ev-Popov-BRST quantization
of three dimensional Chern-Simons gauge theory, with gauge group $G$,
on a three-manifold $M$.
In doing so, in addition to the gauge field $A_a$, one introduces
a ghost field $c$ (anticommuting, of ghost number one, transforming in the
adjoint representation).  The usual BRST transformation laws are
$$ \eqalign{\delta A_a & = -D_a c \cr
                 \delta c & = {1\over 2}[c,c] .\cr} \eqn\overgro$$
We temporarily postpone introducing the antighosts and auxiliary fields
that enter the gauge fixing.

\REF\piguet{O. Piguet and S. P. Sorella, ``On The Finiteness Of BRS Modulo-$d$
Cocycles,'' Univ. of Geneva preprint UGVA-DPT 1992/3-759.}
Just from
the $A-c$ system, it is possible to construct new observables
(discussed, for instance, in [\piguet], where more information can be found).
Let $T$ be an invariant, antisymmetric polynomial on the Lie algebra
${\cal G}$ of $G$.  If $P\in M$ is any point, let
$${\cal O}_T^{(0)}(P) = T(c(P)).            \eqn\nxnx$$
It is evident that ${\cal O}^{(0)}$ is BRST invariant and cannot be written
in the form $\delta(\dots)$.  By solving the ``descent equations''
$$\eqalign{ \d {\cal O}^{(0)} & = \delta {\cal O}^{(1)} \cr
\d {\cal O}_T^{(1)} & = \delta {\cal O}^{(2)} \cr
\d {\cal O}_T^{(2)} & = \delta {\cal O}^{(3)} \cr
\d {\cal O}_T^{(3)} & = 0 , \cr} \eqn\descenteq$$
one finds for each $i$, $0\leq i\leq 3$, an operator-valued $i$-form
${\cal O}_T^{(i)}$ that is BRST invariant up to $\delta (\dots)$.
Hence, if $Y\subset M$ is a $i$-dimensional cycle, then
$$\int_Y{\cal O}_T^{(i)} \eqn\escenteq$$
is BRST invariant (and is easily seen, by virture of \descenteq, to
depend only on the homology class of $Y$).  In particular, setting
$i=3$ and $Y=M$, we get new terms that can be added to the Chern-Simons
Lagrangian.  To be precise, if $L$ is the usual Chern-Simons Lagrangian
and $T_\alpha$ are the antisymmetric invariants on the Lie algebra,
we can take
$$L\to L+\sum_\alpha t_\alpha \int_M\d^3x
\,\,\,  {\cal O}_{T_\alpha}^{(3)}. \eqn\omigo$$
This modification of the standard Chern-Simons theory is implicit in the
work of Kontsevich, who describes the situation in terms of a certain
class of homotopy Lie algebras.

One could also, as in Donaldson theory, pick closed submanifolds
$M_\alpha$ of $M$, of dimension $d_\alpha$, and generalize \omigo\
to
$$L\to L+\sum_\alpha t_\alpha \int_{M_\alpha}\d^3x
 \,\,\, {\cal O}_{T_\alpha}^{(d_\alpha)}. \eqn\tomigo$$

As it stands, \omigo\ is not very useful.
Since the usual Chern-Simons theory conserves ghost number, and
the ${\cal O}^{(3)}$'s all have ghost number $+3$,
the results will be independent of the $t_\alpha$ unless we also introduce
some interaction of negative ghost number.
To get something interesting, we will now modify the usual gauge fixing.

Gauge fixing requires the introduction
of antighost and auxiliary fields.  The standard procedure is to
introduce
the antighost $\overline c$ (anticommuting, of ghost number $-1$,
in the adjoint representation) and the auxiliary field $w$
(commuting, of ghost number 0, in the adjoint representation), with
$$\delta \overline c = i w,~~~~\delta w = 0 . \eqn\vergro$$
Gauge fixing is then carried out by
$$L \to L +\delta \Gamma, \eqn\ergro$$
with any convenient $\Gamma$.  A standard choice involves picking
a metric $g_{ab}$ on $M$.  Writing also $A=A_{(0)}+B$,
where $A_{(0)}$ is a solution of the classical equations about which
one wishes to expand, and denoting the covariant derivative with respect
to $A_{(0)}$ as $D^{(0)}$, we take
$$\Gamma=-\int_M\d^3x\,\,\sqrt g g^{ab}\Tr\overline c D^{(0)}_a B_b.
\eqn\rufo$$
Computing $\delta\Gamma$, one gets the usual gauge fixing Lagrangian
$$ L_{GF}=\int_M\d^3x\,\,\sqrt g\Tr\left(iwD_b^{(0)}B^b-\overline c D_b^{(0)}
D^b c\right). \eqn\dufo$$

Just as in our field theoretic discussion, to modify this, we will consider
not a fixed target $M$ but a family of $M$'s with a variable metric.
Instead of considering the metric $g_{ab}$ of $M$ to be ``inert''
under the BRST transformations, we introduce corresponding fermi variables
$\zeta_{ab}$ (of ghost number 1), with
$$\delta g_{ab}=\zeta_{ab},\,\,\,~~~~\delta\zeta_{ab}=0 .
\eqn\monondd$$
Then \dufo\ is replaced by
$\widetilde L_{GF}=L_{GF} +\Delta L_{GF}$ with
$$\Delta L_{GF}=
-\int_M\d^3x\,\,\sqrt g \left(\zeta^{ab}-{1\over 2}g^{ab}\zeta^c{}
_c\right)\Tr\overline c D^{(0)}_a B_b.
\eqn\rufo$$
Since $\Delta L_{GF}$ is of ghost number $-1$ in the matter fields,
insertions of ${\cal O}^{(3)}$'s can be balanced by insertions of
$\Delta L_{GF}$.  As $\Delta L_{GF}$ is linear in $\zeta$, the
resulting amplitudes, just as in our string theoretic discussion,
will be naturally not numbers but  differential forms
on the  base space of a fibration.

\section{General Target Spaces}

So far we have only considered the $\A$ model with target space
$X=T^*M$.  Now we want to generalize the discussion to consider
an arbitrary symplectic target manifold $X$ (of $c_1=0$),
with $M$ as a Lagrangian submanifold.

The first consequence of replacing
$T^*M$ with a more general $X $ is that there may be nonconstant instantons.
The same argument that we used in proving the vanishing theorem for
$T^*M$ shows that a nonconstant instanton would necessarily have a positive
value of the instanton number
$$q = \int_\Sigma \Phi^*(\omega).     \eqn\jolly$$
To improve the convergence of our formulas, we pick a positive number
$\theta$, and weight instantons of instanton
number $q$ with a factor of $\exp(-\theta q)$.  This can be naturally
built into the formulas by adding to the Lagrangian a suitable multiple of
$q$:
$$L\to L+\theta\int_\Sigma \Phi^*(\omega).   \eqn\olly$$

In the absence of non-constant instantons, the space-time theory
of the $\A$ model was ordinary Chern-Simons theory.  We want to determine
the corrections to this coming from the non-constant instantons.  In doing
so, our goal is to find the classical Lagrangian underlying the space-time
physics of the $\A$ model.  To this end, we concentrate on the case that
the world-sheet is a disc $\Sigma$.  (However, corrections due to
higher topologies can be described similarly.)
The target space Lagrangian $L_T$ is equal to a world-sheet path integral
on the disc:
$$L_T=\int DX\dots D\psi \,\,\,\exp(-L) \eqn\umbo$$
We already know that the contribution to \umbo\ of instanton number $q=0$
is the ordinary Chern-Simons action.  We want to determine the contribution
for some non-zero value of $q$.

Consider  the moduli space of holomorphic maps $\Phi:\Sigma\to X$
with $\Phi(\partial\Sigma)\subset M$ and with two such maps
identified if they differ by an $SL(2,\IR)$ transformation.
The fact that $c_1(X)=0$ and $\dim_\IC(X)=3$ means that in the moduli
problem, the dimensions of the appropriate $H^0$ and $H^1$ are zero.
``Generically'' this means that there are only finitely
many such instantons for each value of $q$.  For simplicity, we will
consider only this case.
If $\Phi$ is such an instanton, let $C=\Phi(\partial \Sigma)$.
Generically, $C$ is a knot in the three-manifold $M$.
Let us work out the contribution of $\Phi$ to the path integral
\umbo.  If we are expanding around a background connection $A$ in
$M$, the contribution of the Chan-Paton factors is
$$ \Tr P\exp\int_C A. \eqn\melbo$$
Note that though $C$ bounds a disc in $X$, it may not do so in $M$,
so \melbo\ can be non-trivial even if $A$ is flat.  We also get
a factor of $\exp(-\theta q)$ from the instanton-counting term in
\olly.  The remaining contributions are nearly trivial since
(i) they are independent of $t$; (ii) they reduce in the large $t$
limit to a ratio of determinants; (iii) except for a possible sign,
the boson and fermion determinants cancel because of the $Q$ symmetry.
The contribution of an instanton is hence
$$ \eta \exp(-\theta q)\Tr P\exp\int_CA     \eqn\elbo$$
where $\eta=\pm 1$ is the ratio of determinants.

The total action is therefore easy to evaluate.  If
$\Phi_i, \,\,\,i=1,2,3,\dots$
are the instantons
of non-zero instanton number, with instanton numbers $q_i$, boundaries
$C_i$, and determinant factors $\eta_i$, then the action is
$$L_T={1\over 2}\int_M\Tr\left(A\wedge \d A+{2\over 3}A\wedge A\wedge A\right)
+\sum_{i=1}^\infty \eta_i\exp(-\theta q_i)\Tr P\exp\int_CA.\eqn\felbo$$

For instance, for $\theta>>0$,
the factors $\exp(-\theta q_i)$ are small, and the instanton corrections
to amplitudes can be evaluated perturbatively.  Their evaluation would
involve calculating expectation values of products of Wilson lines
on the three-manifold $M$.

\section{The $\B$ Model}

Now, we would like, in a similar spirit, to identify the space-time
field theory that is equivalent to the $\B$ model, with the  ``free''
boundary conditions of \S3.1.  To be more precise, we consider
the open string sector of the $\B$ model, and we use the same general
framework of open string field theory as in \S4.1.  We can be brief,
as the arguments are so similar.

As in the case of the $\A$ model, the main simplification comes from
the invariance under rescaling the metric of the target space $X$
by an arbitrary factor $t$.
In \S3.3, we saw that the low-lying modes of the string are
functions ${\cal A}(\phi^I,\eta^{\overline i})$ of the zero modes.
As we now wish ${\cal A}$ to have ghost number 1, we take it to be linear
in $\eta$, so in fact
$${\cal A}=\eta^{\bar i}A_{\bar i}(\phi^I).     \eqn\nucnon$$
So the physical field $A$ is a one-form of type $(0,1)$.
If gauge fields are included via Chan-Paton factors, then $A$ takes
values in $N\times N$ matrices or more generally in the endomorphisms
of some holomorphic vector bundle $E$.

The linearized gauge transformation law $\delta {\cal A}=Q\epsilon$
reduces for large $t$ to $\delta A=\overline\partial \epsilon$,
so $A$ must be interpreted as the $(0,1)$ part of a connection on $E$.
What should be the field equation for $A$?
This can be anticipated from the discussion at the end of \S3.2,
where we showed that the background connections to which the $\B$ model
can be coupled are precisely those for which the $(0,2)$ part of the curvature
vanishes, in other words those that define holomorphic structures on $E$.
To write a Lagrangian from which this equation can be derived, let $\lambda$
be an everywhere non-zero holomorphic three-form on $X$.  Then up to an
undetermined constant, the Lagrangian whose solutions are connections
of vanishing $(0,2)$ curvature is
$$L_T={1\over 2}\int_X\lambda
\wedge \Tr\left(A\wedge \overline\partial A+{2\over 3}
A\wedge A\wedge A\right).            \eqn\ocovo$$
Arguments similar to those that we have given for the $\A$ model
show that in the large $t$ limit, the open string field theory of the
$\B$ model reduces to \ocovo.

The quantum field theory with Lagrangian \ocovo\ is unrenormalizable by power
counting.  However, it has the following all-but-unique property:
there are no possible counterterms that respect the classical symmetries.
The symmetries of \ocovo\ include {\it complex} gauge transformations,
$\overline\partial_A\to g\overline\partial_A g^{-1}$, with $g$ an arbitrary
gauge transformation of $E$ not respecting any reality or unitarity condition;
and local complex changes of coordinates that preserve $\lambda$.  There is
no local density constructed from $A$ that is invariant under these
symmetries.  (Even $L_T$ itself,
though possessing these invariances at least if one considers only
gauge transformations
that are connected to the identity, is not the integral of an invariant
local density.)  Infinities in quantum field theory are ordinarily integrals
of invariant local densities, so if \ocovo\ could be quantized preserving
the symmetries, one would expect this theory to be {\it finite}, though
superficially unrenormalizable.

Relying only on
usual field theory arguments, it is not at all clear that \ocovo\ can
be quantized preserving its symmetries.  However, the equivalence of
\ocovo\ to a string theory strongly suggests that it in fact is finite.
One might worry about whether closed string poles can ruin the finiteness;
at the end of \S5 we will argue that this does not occur.
One might also wonder whether the finite theory given by the string
theory is really \ocovo\ or some more elaborate theory with \ocovo\
coupled to closed strings.  In \S5 we argue that the closed strings
are decoupled.

\chapter{The Closed String Sector}

We have discovered that -- with some reasonable boundary conditions --
the open string sector of the topological $\A$ and $\B$ string theories
has an elegant interpretation in terms of a space-time field theory.
The extension to closed strings does not work so nicely; the first
point of this section is to sketch what the problem is.  After doing this,
I will conclude by trying to show that open and closed strings are
decoupled in these models; this is intended as partial justification for
studying the open strings separately in \S4.

%%%%%%%%%%%%%%%%%%%%%%%%%%%%%%%%%
%Insert Figure 8  csg8X-scale.eps
\epsfxsize=3in
\topinsert\centerline{\epsfbox{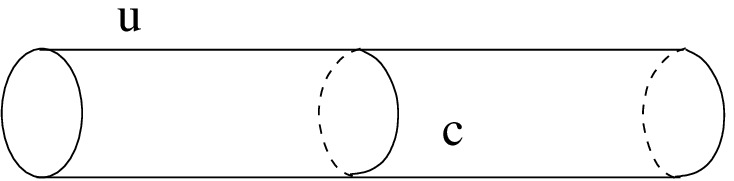}}
{\narrower\smallskip\singlespace
\noindent Fig. 8.
A cylinder $U$ and an embedded circle $C$.  Any diagram
containing
this embedded cylinder has two moduli naturally associated with the cylinder;
one is the length of the cylinder and the second comes from the possibility
of ``cutting'' the cylinder on $C$, and twisting about
$C$ before regluing.
\smallskip} \endinsert
\noindent

\FIG\cylinder{}
%\FIG\cylinder{A cylinder $U$ and an embedded circle $C$.  Any diagram
%containing
%this embedded cylinder has two moduli naturally associated with the cylinder;
%one is the length of the cylinder and the second comes from the possibility
%of ``cutting'' the cylinder on $C$, and twisting about
%$C$ before regluing.}
For open strings, the propagator is
$${b_0\over L_0},\eqn\ncuc$$
with $L_0$ the Hamiltonian of the string and $b_0$ the antighost zero mode.
For closed strings, one has separate zero modes $b_0$ and $\overline b_0$
for right- and left-moving antighosts.  It is convenient to set $b_0^{\pm}
=b_0\pm\overline b_0$.  The formula analogous to \ncuc\ is that the closed
string propagator is
$$ {b_0^-\overline b_0{}^+\over 2L^+_0}\Pi
={b_0\overline b_0\over L^+_0}\Pi.\eqn\jdjdj$$
Here $L^+_0=L_0-\overline L_0$ is the total string Hamiltonian,
and $\Pi$ is the projection operator on states
invariant under rotation of the circle.
This formula is fairly well known, and in any case can be derived
similarly to \ncuc, replacing the strip in figure (\strip) with the cylinder
of figure (\cylinder);
the extra ghost field in the numerator and the projection on rotation-invariant
states come from the twist symmetry of
the cylinder, indicated in the figure.

For the $\A$ model, for instance, the long wavelength limits of $b_0$ and
$\overline b_0$ are the $\partial^*$ and $\overline\partial{}^*$ operators
of the target space $X$.  With $L_0$ reducing at long wavelengths to
the Laplacian $\Delta$, the string propagator looks like
$${\partial^*\overline\partial{}^*\over \Delta}.\eqn\imi$$
This propagator, however, does not seem to arise by gauge fixing of
any local Lagrangian.  It is a pseudodifferential operator of degree zero,
so for it to arise as the inverse of a differential operator, that operator
would have to be of degree zero, that is a constant.

In fact, by repeating for the closed string the analysis of \S3.3,
one finds that the low energy
modes of the closed string sector of the $\A$ model are naturally
represented by a two-form $h$ in space-time.\foot{For the $\B$ model,
one gets instead a sum of $(0,i)$ forms with values in $\wedge^jT^{(1,0)}$,
with $i+j=2$.}
(For large $t$, winding sectors of the closed string cannot
have small eigenvalues of $L_0$.)
The free Lagrangian for
$h$ that one might guess by analogy with our open string results would
be
$$ L=\int_X h\wedge\partial\overline\partial h. \eqn\mcmcxz$$
This has
$${\partial^*\overline\partial{}^*\over \Delta^2} \eqn\cmcmxz$$
for a gauge fixed propagator.  The extra factor of $\Delta $ in the
denominator,
which is in sharp variance with
\imi, of course, makes \cmcmxz\ a pseudodifferential operator of degree $-2$,
in keeping with the fact that the kinetic operator in \mcmcxz\ is second
order.

What sort of Lagrangian do we get by taking the low energy limit
of closed string field theory?
As we discussed in \S2 in a related context,
the free part of the closed string Lagrangian is
$$(\Psi,c_0^-Q\Psi), \eqn\bbxxbxbx$$
where ideally
$$\{b_0{}^-,c_0{}^-\}=1.     \eqn\xbcxbc$$
Such a $c_0{}^-$ does not exist, since $b_0{}^-$, whose field theory
limit is $\partial^*-\overline\partial{}^*$, has a non-trivial cohomology.
However, one can pick a $c_0{}^-$ such that $\{b_0{}^-,c_0{}^-\}=1-T$,
where $T$ is the projection operator onto a subspace (say the kernel of
$L_0$) annihilated by $Q$; this is good enough to ensure gauge invariance
of \xbcxbc.  Such a $c_0{}^-$ is, in the field theory limit,
$${\partial-\overline \partial\over 2\Delta}. \eqn\cucu$$
With this choice of $c_0{}^-$, and recalling that the field theory
limit of $Q$ is $\partial +\overline\partial$,
the field theory limit of the Lagrangian is
not \mcmcxz\ but
$$\int_X h\wedge {1\over \Delta}\partial\overline\partial h. \eqn\ncncoo$$
When this is gauge-fixed and inverted to get a propagator,
the factor of $\Delta^{-1}$
migrates to the numerator, canceling a factor of $\Delta^{-1}$
in \cmcmxz\ and
reproducing the field theory limit of the closed string propagator \imi.

\subsection{A General Puzzle}

The conclusion seems to be that closed string field theory of the $\A$ model
-- or similarly of the $\B$ model -- would be non-local in space-time.
Sometimes such apparent non-localities can be eliminated by introducing
additional fields (such that the apparent non-locality arises in integrating
them out).  I have no evidence that that can be done here.
In any event, certain puzzling  arguments seem to show that the
closed string $\A$ and $\B$ theories do not behave as one would expect
of space-time field theories.

In either the $\A$ or the $\B$ model, there are non-trivial cubic and
higher order couplings of the physical modes.
In studying the open string sector of the $\A$ model, there was
at the classical level (the world-sheet being a disc)
a cubic coupling of three physical fields given by the classical
formula $\int_M\Tr A\wedge A\wedge A$, and various higher interactions
involving Massey products, as we saw in \S4.2.  As always in
field theory, these couplings are Taylor series
coefficients of a natural potential
$V(t_i)$ for sources $t_i$
coupling to the physical modes.  The classical
solutions of the space-time theory -- or equivalently the possible
world-sheet theories --
are in one-to-one correspondence with the critical points of $V$.

What about the closed string sectors?  There are analogous cubic and
higher couplings of physical modes.  For instance, for the closed string
sector of the $\A$ model, the cubic coupling in the large volume limit
is $\int_X h\wedge h \wedge h$.  Experience with both field theory and
with the open string sectors of the $\A$ and $\B$ models
lead us to form a generating function
$V(t_i)$ from these couplings, with the expectation that the allowed
world-sheet theories will correspond to the critical points of $V$.
This latter expectation proves to be false.   For the $\A$ model,
for instance, the part of the $h$ field that is a $(1,1)$ form in space-time
represents a displacement in the Kahler class of the metric of $X$.
The $\A$ model makes sense for any choice of this Kahler class,
so in contrast to what one would have anticipated,
having $V'(t_i)\not= 0$ is not an obstruction to being able to define
the world-sheet theory.
Likewise, for the $\B$ model, the low energy modes include a displacement
in the complex structure of $X$.  The $\B$ model makes sense for
any complex structure on $X$, even though there are non-zero cubic
and higher order couplings for the fields representing a displacement in
the complex structure.

\section{Closed String Contributions To Amplitudes}

Leaving this puzzle as food for thought, I will conclude by making
some simple comments about the closed string contributions to
open string amplitudes of the $\A$ and $\B$ models.
%%%%%%%%%%%%%%%%%%%%%%%%%%%%%%%%%
%Insert Figure 9  csg9Xa-b.eps
\epsfxsize=3in
\topinsert\centerline{\epsfbox{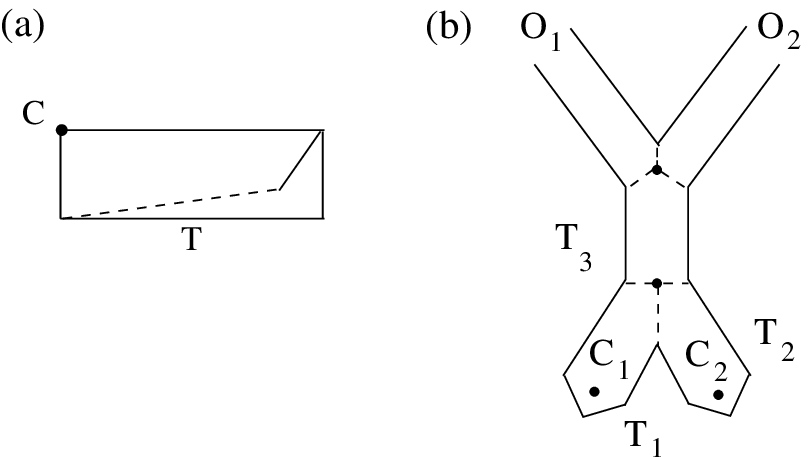}}
{\narrower\smallskip\singlespace
\noindent Fig. 9.
In the fatgraph description of a Riemann surface coupling open and closed
strings, the closed strings are incorporated as in (a).  A standard
open string strip of width $\pi$ and length $T$,
open at one end (where it attaches to the rest of
the diagram) is closed at the other end by folding it over on itself;
the closed string is attached at the resulting conical singularity,
as shown.
A typical fatgraph coupling open and closed strings is shown in (b).
It consists of five flat strips glued on the dotted lines, with conical
singularities shown as solid dots.
The open strings $O_1$ and $O_2$ are attached at the ends of infinite
flat strips, while the closed strings appear as conical singularities
on strips of finite lengths $T_1$ and $T_2$.  This particular world-sheet
has three real moduli: $T_1$, $T_2$, and the length $T_3$ of the one
internal propagator.
\smallskip} \endinsert
\noindent

\FIG\fatcoup{}
%\FIG\fatcoup{
%In the fatgraph description of a Riemann surface coupling open and closed
%strings, the closed strings are incorporated as in (a).  A standard
%open string strip of width $\pi$ and length $T$,
%open at one end (where it attaches to the rest of
%the diagram) is closed at the other end by folding it over on itself;
%the closed string is attached at the resulting conical singularity,
%as shown.
%A typical fatgraph coupling open and closed strings is shown in (b).
%It consists of five flat strips glued on the dotted lines, with conical
%singularities shown as solid dots.
%The open strings $O_1$ and $O_2$ are attached at the ends of infinite
%flat strips, while the closed strings appear as conical singularities
%on strips of finite lengths $T_1$ and $T_2$.  This particular world-sheet
%has three real moduli: $T_1$, $T_2$, and the length $T_3$ of the one
%internal propagator.}

\REF\otherthorn{D. Z. Freedman, S. B. Giddings, J. A. Shapiro, and C. B.
Thorn, Nucl. Phys. {\bf B298} (1988) 253.}
Let us look briefly at the couplings of open and closed strings in
these models.
To do so we will use the fatgraph or open string field theory description
of external open and closed strings.
An external open string (a marked point on the boundary of a Riemann surface)
should have one real modulus, while an external closed string (a marked point
in the interior) should have two.
In the fatgraph description, as one might expect, external open strings
are represented by open string propagators going off to infinity,
as for $O_1$ and $O_2$ in figure (\fatcoup(b)).  The one real modulus
of the open string is, roughly, the location at which its propagator is
attached to the rest of the figure.  The proper coupling of external
closed strings (which can be deduced by seeing how closed string
poles arise in open string diagrams, as in [\otherthorn]) is as follows.
One attaches to the open string diagram an external open string propagator
of {\it finite} length $T$, closes it up by folding together its free
end, and inserts the closed string at the resulting conical singularity.
The two real moduli of the closed string are, roughly, $T$ and
the position at which the propagator is attached to the rest
of the diagram.
This is sketched in figures (\fatcoup(a,b)).

To compute the classical couplings of $n$ open strings and $m$ closed
strings, we consider the moduli space ${\cal D}_{n,m}$ of a disc
with $n$ marked points on the boundary and $m$ in the interior;
its real dimension is $n+m-3$.  (The analysis goes through the same way
in higher genus.)
Comparing the dimension of ${\cal D}_{n,m}$ to that
of ${\cal D}_{n+m,0}$, there
is one extra modulus for each marked point in the interior.
As explained in the last paragraph,
the fatgraph description of ${\cal D}_{n,m}$ (figure (\fatcoup(b)))
is similar to the fatgraph
description of ${\cal D}_{n+m,0}$ except that, while the external open
strings are attached to outgoing propagators of infinite length, the external
closed strings are attached to propagators of variable length;
the one extra real modulus for each external closed string is
precisely
the length of the propagator to which it is attached.

In the $\A$ model, for instance, an external open string is represented
by an (${\rm End}
(E)$-valued) one-form $\alpha$, while an external closed string is represented
by a two-form $h$.  Integration over the extra modulus of the
propagator by which the closed string is attached to the rest of the diagram
multiplies $h$ by the {\it open string} propagator $b_0/L_0$.  In the large
$t$ limit, this simply
turns $h$ into a one-form
$$ {\d^*\over L_0}\cdot h.      \eqn\nsnsn$$
This one-form then couples as just one more external open string
state (which happens to be valued in the center of ${\rm End}(E)$).
But the one-form in \nsnsn\ is exact (since, for instance, the open string
propagator annihilates harmonic forms),
so the corresponding open string state
decouples.  Consequently the on-shell couplings
of open and closed strings are all zero -- in either the $\A$ or $\B$
model.

\REF\witn{E. Witten, ``Quantum Field Theory And The Jones Polynomial,''
Comm. Math. Phys. {\bf 121} (1989) 351.}
\REF\raysinger{D. Ray and I. M. Singer, ``Analytic Torsion And The Laplacian
On Complex Manifolds'' Ann. of Math. {\bf 98} (1973) 154.}
If the closed strings are, then, decoupled from the open strings,
what is their role?  The following
conjecture seems natural to me.  The field theories we have extracted
from the $\A$ and $\B$ models all have $c$-number anomalies, analogous
to the central charge in two-dimensional conformal field theory.
For instance, the anomalies of Chern-Simons theory
are connected with framings of three-manifolds and of knots [\witn].
One-loop anomalies of the field theory related to the $\B$ model
were calculated long ago by Ray and Singer [\raysinger].
These anomalies are possible only because of the ultraviolet difficulties
of quantum field theory, which of course are greatly ameliorated in string
theory.  It therefore seems reasonable to suspect that the closed
string contributions in the $\A$ and $\B$ theories cancel the anomalies
of these theories, without, in view of the decoupling argued above,
having much effect on the open string ``physics.''

In any event, whatever the closed string contributions,
they are finite. In usual string theory, possible infinities
come from physical closed string poles, but in these topological
models, there are no such poles since the closed string
propagator annihilates the physical states or
harmonic forms; this is because
(in contrast to conventional string backgrounds)
the $b_0$ operator annihilates the kernel of $L_0$.  Thus, for instance,
in the case of the $\B$ model, whose finiteness perhaps comes as
a surprise (since the field theory related to the low energy limit
of the open strings
is superficially unrenormalizable), the closed strings will not ruin
this finiteness.

\ack{The discussion of closed string field theory in \S2 benefitted
considerably from discussions with B. Zwiebach.  Several relevant
references were pointed out by L. Jeffrey.}

\endpage
\refout
%\figout
\end